\documentclass[aps,prl,twocolumn,amsmath,amssymb,superscriptaddress,nobibnotes,showpacs,floatfix]{revtex4-1}
\usepackage{graphicx}

\usepackage{amsmath}
\usepackage{amssymb}
\usepackage{amsthm}
\usepackage{tabularx}
\usepackage{multirow}

\usepackage[usenames, dvipsnames]{color}

\usepackage{hyperref}
\hypersetup{
  colorlinks = true,    
  allcolors = blue
}

\def\vR{{\bf R}}

\begin{document}

\title{Interplay between multipolar spin interactions, Jahn-Teller effect and electronic correlation in a $J_{eff}=\frac{3}{2}$ insulator}

\author{Dario Fiore Mosca}
\affiliation{University of Vienna, Faculty of Physics and Center for Computational Materials Science, Vienna, Austria}

\author{Leonid V. Pourovskii}
\address{Centre de Physique Th\'eorique, Ecole Polytechnique, CNRS, Institut Polytechnique de Paris, 91128 Palaiseau Cedex, France}
\address{Coll\`ege de France, 11 place Marcelin Berthelot, 75005 Paris, France}

\author{Beom Hyun Kim}
\affiliation{Korea Institute for Advanced Study, Seoul 02455, South Korea}

\author{Peitao Liu}
\affiliation{University of Vienna, Faculty of Physics and Center for Computational Materials Science, Vienna, Austria}

\author{Samuele Sanna}
\affiliation{Department of Physics and Astronomy, Alma Mater Studiorum - Universit\`{a} di Bologna, Bologna, 40127 Italy}

\author{Federico Boscherini}
\affiliation{Department of Physics and Astronomy, Alma Mater Studiorum - Universit\`{a} di Bologna, Bologna, 40127 Italy}

\author{Sergii Khmelevskyi}
\affiliation{Center for Computational Materials Science, Institute for Applied Physics, Vienna University of Technology, Wiedner Hauptstrasse $8$ - $10$, $1040$ Vienna, Austria}

\author{Cesare Franchini}
\email{cesare.franchini@univie.ac.at}
\affiliation{University of Vienna, Faculty of Physics and Center for Computational Materials Science, Vienna, Austria}
\affiliation{Department of Physics and Astronomy, Alma Mater Studiorum - Universit\`{a} di Bologna, Bologna, 40127 Italy}

\date[Dated: ]{\today}

\begin{abstract}
In this work we study the complex entanglement between spin interactions, electron correlation and Janh-Teller structural instabilities in the 5d$^1$ $J_{eff}=\frac{3}{2}$  spin-orbit coupled double perovskite $\rm Ba_2NaOsO_6$ using first principles approaches. By combining non-collinear magnetic calculations with multipolar pseudospin Hamiltonian analysis and many-body techniques we elucidate the origin of the observed quadrupolar canted  antifferomagnetic. We show that the non-collinear magnetic order originates from Jahn-Teller distortions due to the cooperation of Heisenberg exchange, 
quadrupolar spin-spin terms and both dipolar and multipolar Dzyaloshinskii-Moriya interactions.
We find a strong competition between ferromagnetic and antiferromagnetic canted and collinear quadrupolar magnetic phases:
the transition from one magnetic order to another can be controlled 
by the strength of the electronic correlation ($U$) and by the degree of Jahn-Teller distortions.

\end{abstract}
\maketitle

Double perovskites with strong spin-orbit coupling (SOC) represent a unique playground for the emergence of exotic spin-orbital-lattice entangled phases~\cite{PhysRevB.82.174440, PhysRevB.90.184422, PhysRevB.98.075138,PhysRevLett.118.217202, Martins_2017}. Structural and orbital frustrations~\cite{Khomskii_2003}, SOC-induced formation of \emph{effective} J manifolds~\cite{PhysRev.171.466} and relatively strong electronic correlation give rise to a wide range of unusual electronic and magnetic phases including so-called Dirac-Mott insulating states~\cite{Lee_2007, PhysRevB.75.052407}, multipolar spin interactions~\cite{PhysRevB.82.174440, PhysRevB.90.184422, Lee_2007} and entangled SOC-Jahn-Teller effects~\cite{PhysRevB.98.075138, Xu2016}.

\begin{figure}[h]
   \begin{center}
        \includegraphics[width=1.00\columnwidth,clip=true]{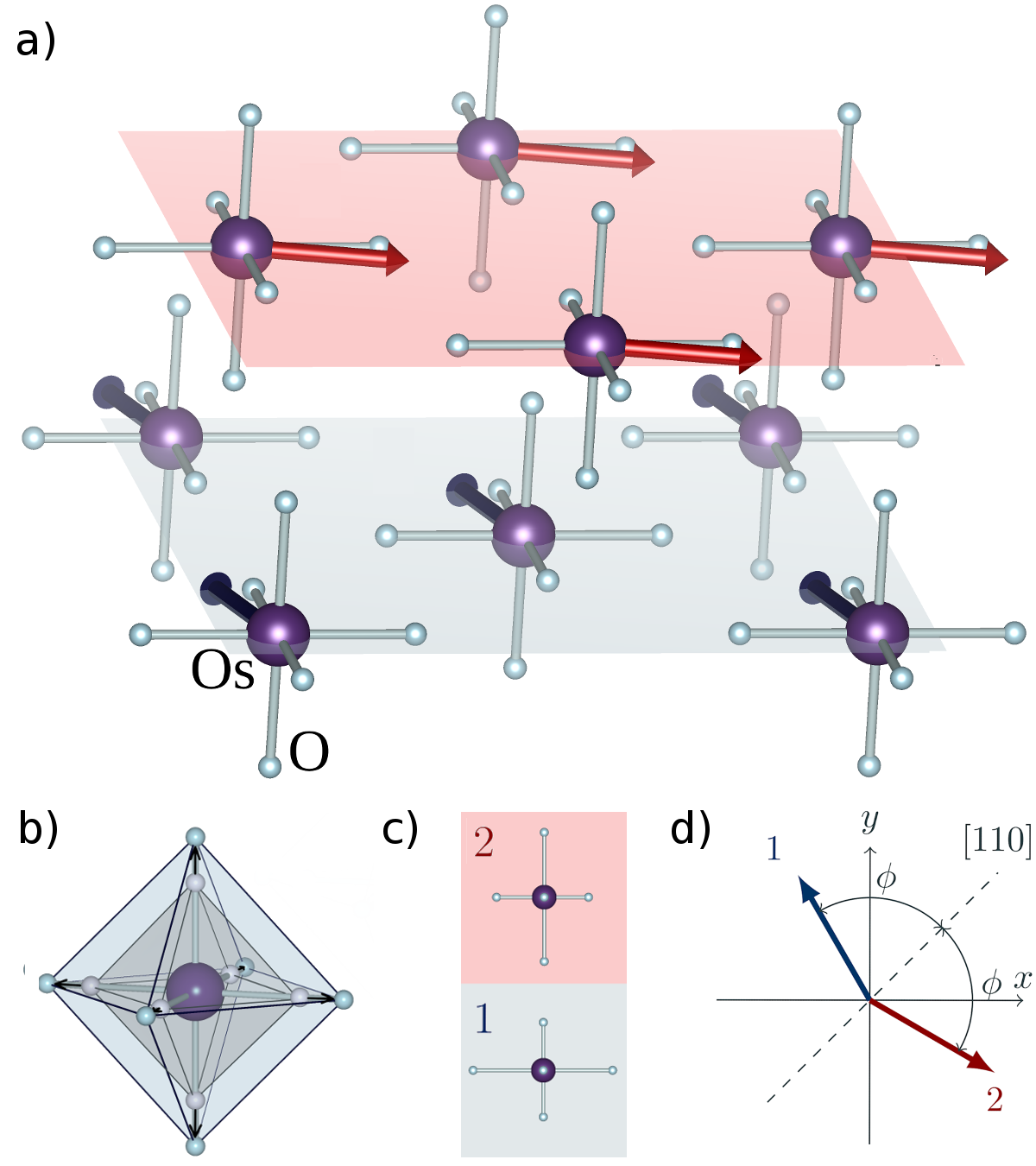}
    \end{center}
\caption{(Color online) Graphical representation of (a) the canted-AFM magnetic ground state with layer-dependent (1 \& 2) spin ordering and (b) associated JT vibrational mode. Following Van Vleck notation~\cite{Vleck1939,supplmat} we obtain $Q_1=0.031~\AA$ and $\theta = \tan^{-1}(Q_2/Q_3) = 79.35^\circ$ (plane 1) and -79.35$^\circ$ (plane 2)~\cite{doi:10.1063/1.1984590}.
This structural motif is compatible with model A proposed by Liu \emph{et al.}~\cite{PhysRevB.97.224103}. (c) Visualization of the staggered-type  JT-distortion in the two layers. (d) Schematic definition of the canting angle $\phi$: $\phi$=0 corresponds to the FM limit; by convention $0 < \phi < \pi/4$ refers to c-FM state whereas  $\pi/4 < \phi < \pi/2$ identifies a c-AFM spin arrangement. Optimized structural data and JT distortions are given in the Supplementary Materials (SM)~\cite{supplmat}.
}
\label{fig:1}
\end{figure}

A particularly interesting case is that of 5$d^1$ double perovskites characterized by an effective total angular momentum $J_{eff}=3/2$, an example of which is $\rm Ba_2NaOsO_6$ (BNOO). 
BNOO is Dirac-Mott insulator~\cite{PhysRevB.94.235158,Lee_2007} with strong SOC and electronic correlation effects~\cite{PhysRevB.91.045133, PhysRevB.93.155126, PhysRevB.99.035126}.
Magnetization measurements indicate that below 6.8~K BNOO exhibits weak ferromagnetic interactions and a small ordered spin moment of 0.2~$\mu_B$~\cite{STITZER2002311,PhysRevLett.99.016404,PhysRevB.84.144416, PhysRevB.95.064416}. Refined NMR analysis confirmed the theoretically predicted canted spin state with two magnetic sublattices~\cite{PhysRevB.90.184422,Lu2017}. In this spin model, depicted in Fig.\ref{fig:1}(a), the spins align ferromagnetically within the $xy$ plane with alternating canting angle $\phi$ in adjacent planes along the [001] axis [see Fig.\ref{fig:1}(c)]. As the optimal canting angle extracted from the NMR data, $\phi_{expt}$=67$^\circ$, is closer to the AFM limit (90$^\circ$) than to the FM one (0$^\circ$) it is more appropriate to use the jargon canted-antiferromagnetic (c-AFM), rather than c-FM as initially proposed~\cite{Lu2017, PhysRevB.90.184422,PhysRevB.100.041108}. The most stunning aspect of BNOO is that this c-AFM phase is associated with a local point symmetry breaking manifested by local deformations of the BO$_6$ octahedra~\cite{Lu2017, PhysRevB.97.224103}. However, x-ray diffraction shows that the cubic symmetry is maintained even at low temperature~\cite{PhysRevLett.99.016404} and no direct evidence of Jahn-Teller (JT) instabilities have been detected by NMR~\cite{Lu2017,PhysRevB.97.224103}, even though JT instabilities would be expected in a $d^1$ electronic configuration~\cite{Khomskii_2003,khombook} (one of the models proposed in Ref.~\cite{Lu2017} involves a contraction of the OsO$_6$ octahedra, assimilable to a local Q$_1$-JT distortion, see Fig.~\ref{fig:1}). 
This apparent violation of the JT theorem~\cite{jt2, Gehring_1975} has been explained by the entanglement of vibronic dynamics with SOC, suggesting the presence of a dynamical Jahn-Teller effect with small static component~\cite{PhysRevB.98.075138, Xu2016} and orbital selective quadrupolar charge ordering~\cite{PhysRevB.100.245141}.
This explains the breaking of local point symmetry and suggests that the canted magnetic ground state of BNOO should be described in terms of the spin-orbital-lattice entangled states involving high-rank spin  interactions. 

With this study we aim to clarify the link between JT instabilities and spin couplings by computing the variation of the multipolar interactions due to local structural distortions and explain how this JT-multipolar coupling affects the spin deflection and dictates the formation of the c-AFM ground state.
We do so by employing first principles non-collinear relativistic density functional theory (DFT)~\cite{PhysRevB.62.11556,Liu2015,PhysRevMaterials.3.083802} within the 
Perdew-Burke-Ernzerhof approximation~\cite{PhysRevLett.77.3865}
using the  VASP code~\cite{PhysRevB.47.558, PhysRevB.54.11169,PhysRevB.50.17953} and many-body techniques.
The inter-site interactions have been extracted from magnetically constrained DFT+U+SOC calculations, from DFT+dynamical mean-field theory (DMFT)~\cite{RevModPhys.68.13,Anisimov_1997} in the Hubbard-I approximation (HI)~\cite{hubbard_1,lichtenstein_dft_dmft} in conjunction with the magnetic force theorem (FT) to effective intersite interactions (abbreviated, FT-HI)~\cite{PhysRevB.94.115117}, as well as from exact diagonalization (ED) of a microscopic effective Hamiltonians~\cite{PhysRevLett.109.167205}. The charge self-consistent DFT+HI approach\cite{Aichhorn2009,Aichhorn2011} is implemented using the Wien-2k package~\cite{Wien2k} and TRIQS library~\cite{Parcollet2015,Aichhorn2016} using the projective technique of Refs.~\cite{Amadon2008,Aichhorn2009}. 
A more detailed description of the computational protocol is given in the method section
and in the SM~\cite{supplmat}.

We find that in the fully undistorted (JT-quenched) limit the spins prefer to align ferromagnetically but the onset of staggered JT distortions activates both first-order (dipole-dipole) and second order (quadrupole-quadrupole + dipole-octupole) Dzyaloshinskii-Moriya (DM) interactions~\cite{Hosoi} which drive the development of the observed c-AFM phase at the optimum JT distortion. Moreover, our data show that the multipolar c-AFM ground state is highly sensitive to the degree of JT distortions and to the strength of the Coulomb interaction $U$: slightly lower $U$ or larger JT distortions shift the system to a competing canted-ferromagnetic (c-FM) state.

\begin{figure*}[t]
   \begin{center}
   \includegraphics[width=1.00\textwidth,clip=true]{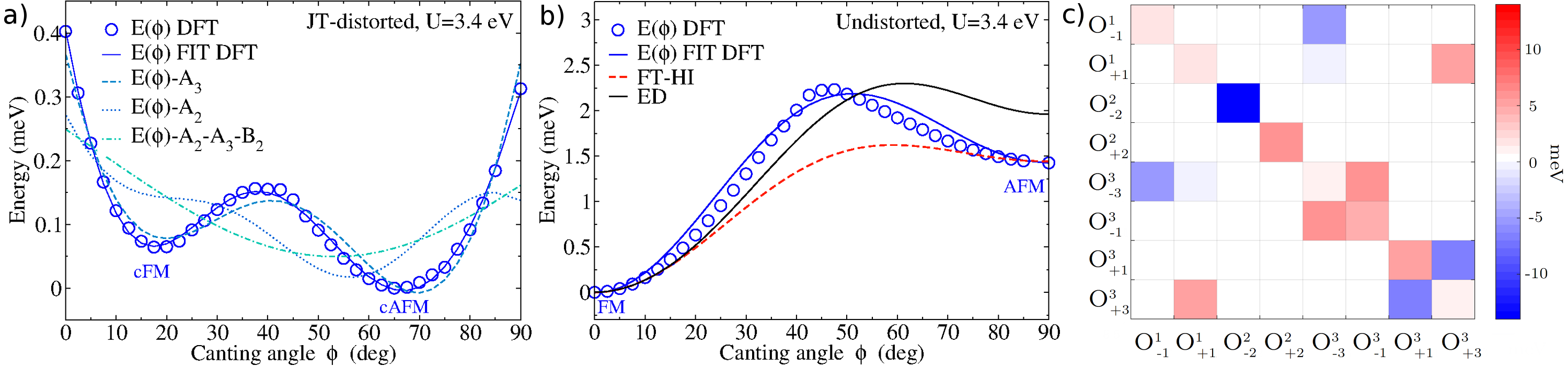}
    \end{center}
\caption{(Color online) Dependence of the energy on the canting angle and spin-spin interaction matrix. (a) Calculated DFT+U+SOC $\phi$ dependent energies for the optimized JT distorted phase (circles) and corresponding fitting curve obtained using Eq.~\ref{eq:fitting}. The fitting curves obtained by removing  the A$_3$ (o-o), the A$_2$ (q-q)  and A$_2$-A$_3$-B$_2$ (q-q- and o-o) coefficients are also displayed; (b) Comparison between DFT, FT-HI and ED canting energies for the undistorted (Q$_1$=0) phase. (c) FT-HI interaction matrix showing the non-zero and non-constant mean values of the tensor operators $\langle{O}\rangle$ in the pseudo spin eigenbasis (see Eq.~\ref{eq:interaction})
summed over the out-of plane interacting bonds (see SM for details~\cite{supplmat}). To guarantee a quantitatively consistent comparison between DFT and FT-HI the DFT FM-AFM energy difference is aligned to that obtained by FT-HI (see SM for details~\cite{supplmat}).}
\label{fig:2}
\end{figure*}

We start by comparing the dependence of the energy on the canting angle for the JT-distorted and JT-quenched (undistorted) phases in Fig.\ref{fig:2}, obtained by changing stepwise the canting angle $\phi$ from the FM ($\phi=0^\circ$) to the AFM ($\phi=90^\circ$) limits. The JT structure exhibits two minima, a lower c-AFM ground state found at $\phi=66^\circ$, in excellent agreement with experiment (67$^\circ$\cite{Lu2017}) and a c-FM phase just 0.05 meV/u.c. higher in energy.
Conversely, the undistorted structure shows a clear minimum for a parallel spin alignment followed by an AFM state. The size of the spin ($m_s$) and orbital ($m_o$) moments are essentially the same in both structures ($m_s\approx0.8~\mu_B$ and $m_o\approx0.6\mu_B$). In the ground state JT-distorted structure the resulting net ordered moment is $0.23\mu_B$, in very good agreement with measurements (0.2$\mu_B$)~\cite{PhysRevLett.99.016404, Lu2017}. These results clearly indicate that the JT effect is essential to stabilize the observed canted ground state. 

To decipher the role of JT distortions on the spin deflection we have extracted the intersite spin interactions by fitting the canting energies curves with a pseudospin Hamiltonian in the general multipolar form~\cite{ARIMA1969, RevModPhys.81.807}:

\begin{equation}
H_{ij} =\sum_{K,K'}\sum_{Q,Q'}I_{KK'}^{QQ'}O_{Q}^{K}(J_i)O_{Q'}^{K'}(J_j),
\label{eq:interaction}
\end{equation}
where  $i$,$j$ are the site indexes, $I_{KK'}^{QQ'}$ the coupling constants and  $O_{Q}^{K}(J_i)$ and $O_{Q'}^{K'}(J_j)$ are the
multipolar tensor operators of rank $K$ and $K'$,  
with $K \le 2J_i$, $Q = -K, ..., K$, $K' \le 2J_j$, $Q' = -K',..., K'$.
The multipolar operators $O$ are estimated within the mean field approximation, i.e. $O(J_i)O(J_j) \approx \langle{O(J_i)}\rangle{O(J_j)} + O(J _i)\langle{O(J _j)}\rangle 
- \langle{O(J_i)}\rangle \langle{O(J_j)}\rangle$. After applying symmetry properties of pseudovectors and multipolar operators for the $J_{eff}=\frac{3}{2}$ (or $\Gamma_8$) state~\cite{doi:10.1143/JPSJ.67.941,PhysRevB.82.174440} as well as  crystal symmetry we obtained the following fitting formula which expresses the magnetic energy as a function of the canting angle $\phi$ (the derivation is given in the SM~\cite{supplmat}):

\begin{eqnarray}
E(\theta) & = & A_1 \cos(2\phi) + A_2 \cos(4\phi) + A_3 \cos(6\phi) \nonumber \\
          &   & + B_1 \sin(2\phi)+ B_2 \sin(4\phi)
          \label{eq:fitting}
\end{eqnarray}

To validate the DFT-mapping and decode the specific type of spin couplings at play
we have performed a direct calculation of the intersite interactions using 
the FT-HI method proposed by Pourovskii~\cite{PhysRevB.94.115117} and ED~\cite{PhysRevLett.109.167205}, taking the undistorted phase as a reference (FT-HI and ED cannot be applied to the JT-distorted  phase). 
All possible on-site  fluctuations within the paramagnetic ground state of a shell, encoded by the corresponding multipole moments $O_{Q}^{K}(J_i)$,   are included; all relevant inter-site couplings can thus be extracted. In the present case of the $J_{eff}=\frac{3}{2}$ shell of Os, those are all inter-site interactions between dipole, quadrupole, and octupole moments that are permitted by the symmetry. With the effective Hamiltonian thus obtained we evaluated the dependence of zero-temperature total energy versus $\phi$;  the spin-spin coupling parameters $A_i$ are subsequently extracted by fitting it to the form (\ref{eq:fitting}).

\begin{table}[b]
\centering
\begin{tabular}{ll|c|c|c|c}
\hline
\hline
\multicolumn{6}{c}{ Spin-spin Interactions}\\
\hline
\hline
  &      & \multicolumn{3}{c|}{Undistorted} & {JT-distorted}\\
  &  & \multicolumn{3}{c|}{cFM} & {cAFM} \\
\hline
\multicolumn{2}{l|}{Coefficients}  &   FT-HI   & ED    & DFT     & DFT \\\hline
A$_1$        &                     &  -0.72  & -1.06 & -0.66   & 0.13       \\
             &d-d                  &   0.84  &  0.39 &         &     \\
             &q-q                  &  -0.03  &  0.00 &         &  \\
             &d-o                  &  -1.32  & -1.25 &         & \\                    
             &o-o                  &  -0.21  & -0.20 &         & \\
A$_2$        &q-q                  &  -0.40  & -0.48 & -0.69   & -0.38\\
A$_3$        & o-o                 &  0.05   & 0.08  & -0.03   &-0.08 \\
B$_1$        & DM d-d              &  0      & 0     &    0    & -0.99\\
B$_2$        & DM q-q, d-o         &  0      & 0     &    0    &-0.08\\
\hline
\end{tabular}
\caption{Collection of spin-spin interactions (in meV) as derived from DFT, FT-HI and ED approaches for the undistorted Q$_1$=0 FM phase and corresponding DFT values for the fully distorted c-AFM phase. The coefficients are decomposed over their distinct d-d, d-o, o-o, and q-q character.}
\label{tab:coefficients}
\end{table}

The fitting curves follow very well the first principles data for both the distorted and undistorted phases as shown in Fig.~\ref{fig:2}(a,b). 
The optimal fitting parameters and the corresponding FT-HI and ED results are collected in Tab.~\ref{tab:coefficients} where each coefficient is associated with the specific type of dipolar or multipolar couplings.  Both phases exhibit relatively large quadrupole-quadurpole (q-q) couplings and non negligible octupole-octupole (o-o) interactions.
The inclusion of o-o terms in the JT-phase is essential to shift the minimum from 70$^\circ$ to 66$^\circ$, whereas the  formation of the two competing canted minima is mostly governed by the q-q term $A_2$.  We also note that DM interactions, embedded in coefficients $B_1$ and $B_2$, are inoperative in the undistorted phase (not allowed by symmetry), but the breaking of the local symmetry caused by the JT effect activates the DM channel which is the main responsible for stabilizing of the canted phases. Our data indicate that, besides the standard first order d-d DM interaction, an additional second order DM coupling emerges involving both q-q and d-o terms. This type of interaction can be classified as multipolar DM interaction, namely an antisymmetric DM exchange among multipolar tensors. 

\begin{figure*}[t]
   \begin{center}
   \includegraphics[width=1.0\textwidth,clip=true]{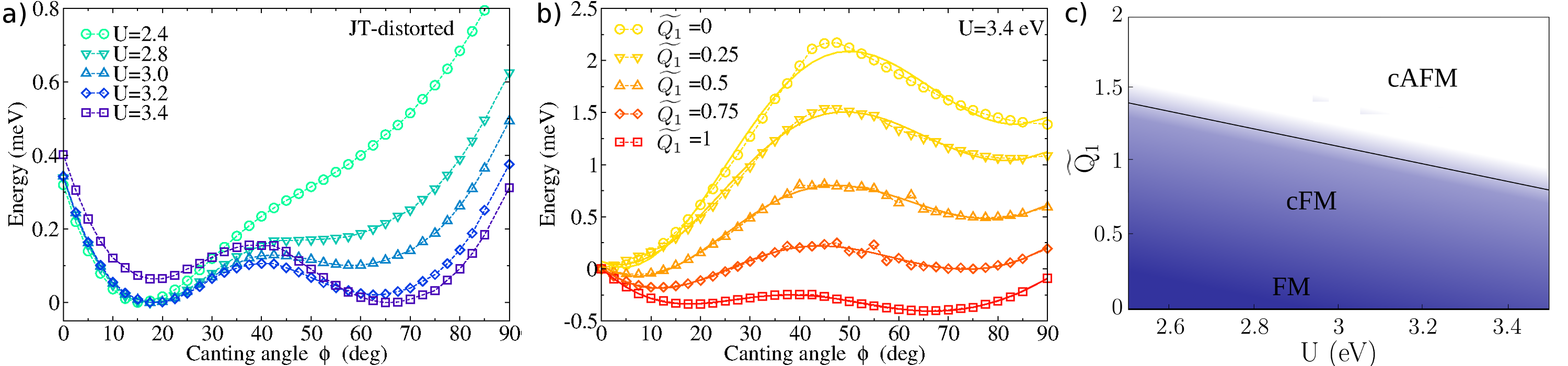}
    \end{center}
\caption{(Color online) Magnetic canting energy as a function of (a) $U$ (relative global minimum set to zero energy) and (b) $\widetilde{Q_1}$ ($\widetilde{Q_1}=1$ refers to the fully optimized JT phase). (c) Global magnetic phase diagram showing the dependence of the ground state magnetic ordering (canting angle) for different values of $U$ and $\widetilde{Q_1}$. Decreasing $\widetilde{Q_1}$ drives the system to a collinear FM state whereas a reduction of $U$ washes-out the c-AFM system and stabilizes a c-FM configuration.  
}
\label{fig:3}
\end{figure*}

After clarifying the essential features of the c-AFM ground state we move now to the analysis of the competition between the c-AFM and c-FM phases and how this competition is controlled by the strength of $U$ and the degree of JT mode $Q_1$. Being almost degenerate in energy, it is expected that small perturbations could change the relative stability of these two phases, providing insights on the mechanism favoring one or the other spin ordering. To explore this possibility we have computed the canting energies for different values of the relative JT distortion $\widetilde{Q_1}$ (from the fully distorted structure, here identified as $\widetilde{Q_1}=1$ to the undistorted $\widetilde{Q_1}$=0 limit) and $U$ (from 3.4 to 2.4 eV). The results, displayed in Fig.~\ref{fig:3}, indicate that both degrees of freedom act on the spin deflection and can trigger a magnetic transition from the c-AFM to c-FM.

By decreasing $U$ the c-AFM phase is progressively destabilized in favor of the c-FM phase which becomes the global minimum. A further decrease of $U$ washes-out completely the AFM canting. Similarly, reducing $\widetilde{Q_1}$ acts on the canting angle by inverting the stability of the two canted minima and drives the systems towards the collinear limits as $\widetilde{Q_1}$ approaches 0. In order to explore the combined action of JT effect and electronic correlation we have repeated similar calculations and constructed a complete magnetic phase diagram as a function of $\widetilde{Q_1}$ and $U$, displayed in Fig.~\ref{fig:3}(c). The phase diagram shows three different phases characterized by specific value of the canting angle: (i) a FM phase which is favored in the limit of vanishing JT-distortion and small $U$; (ii) a c-FM phase emerges beyond a small critical JT distortion in the whole considered $U$-regime and finally (iii) the c-AFM phase, corresponding to the experimentally observed ground state, which covers the top portion of the phase diagram and is favored by increasing the JT mode and $U$. For $\widetilde{Q_1}$ =1 the c-AFM phase represents the ground state for $U$ larger than 3.1 eV.

\begin{figure}[h]
   \begin{center}
   \includegraphics[width=0.49\textwidth,clip=true]{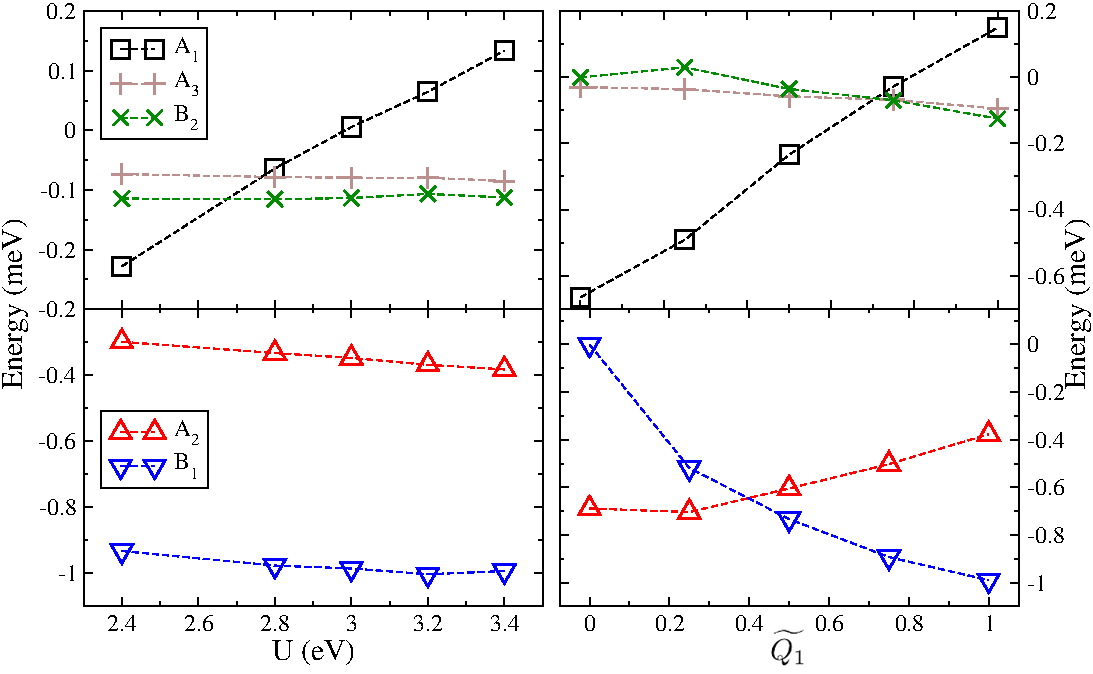}
    \end{center}
\caption{(Color online) Evolution of the DFT+U spin-spin exchange coefficients as a function of (left) $U$ and (right) $\widetilde{Q_1}$.}
\label{fig:4}
\end{figure}

The interplay between $\widetilde{Q_1}$ and $U$  affects the exchange couplings. The transition from one phase to the other can be rationalized by inspecting the variation of the spin-spin interactions upon $U$ and $\widetilde{Q_1}$ shown in Fig.~\ref{fig:4}. First we note that the tendency to move toward a FM-type canting is mostly triggered by the steep decrease of the d-d interaction embedded in the exchange coefficient $A_1$, which changes sign rather quickly when $U$ or $\widetilde{Q_1}$ are reduced, thereby driving the AFM-to-FM transition. We note that coefficients $A_3$ and $B_2$ besides being relatively small remain essentially unaffected by tuning $U$ at the optimized JT distorted structure,
implying that the o-o character is mostly linked to the structural instabilities. 
This is confirmed by the observation that $A_3$ is rapidly diminished with decreasing $\widetilde{Q_1}$ thus revealing a direct link between o-o interaction and structural distortion.
The JT distortion is therefore a determinant factor for the onset of the dipolar and multipolar DM interactions expressed by the coefficient $B_1$ and $B_2$, respectively. Similarly to the multipolar DM component also the dipolar one  ($B_1$), which is zero by symmetry in the undistorted limit, rapidly increases vs. $\widetilde{Q_1}$ and reaches a large value in the  fully  JT-distorted phase ($B_1=-0.99$~meV). Finally, from the trend of $A_2$ we infer that q-q interactions become stronger with decreasing $\widetilde{Q_1}$ and are responsible for the preservation of the two minima  in the undistorted phase. Conversely, reducing $U$ leads to a linear reduction of $A_2$, which contributes to the disruption of the large-$\phi$ minimum.
{{The above analysis can give insights on the role of dynamical JT effects in BNOO, that should be manifested by two types of magnetic fluctuations: FM fluctuations, controlled by the changes in $A_1$, and canted-FM to canted-AFM fluctuations governed by the variation of  $B_1$.}}

In conclusion, by combining many-body effective Hamiltonians with material specific DFT-based descriptions we have decoded the complex entanglement between structural frustration and electronic correlation in the spin-orbit coupled and spin-canted   double perovskite BNOO. 
We show that this exotic magnetic order originates from the anomalous Dzyaloshinskii-Morya interactions generated by JT -distortions,
and elucidate the direct effect of U and JT on the spin-spin interactions. This work contributes to the raising interest on atypical multipolar orderings in relativistic oxides~\cite{PhysRevLett.124.087206,PhysRevB.101.054439} with a transparent analysis that helps to light up the intricate physics at play in this class of materials. 

\section{Methods}
To study the cross coupling between spin exchange interaction, Jahn-Teller distortions and electronic correlation we have integrated three different schemes:
magnetically constrained non-collinear DFT+U, DMFT in the Hubbard-I approximation (FT-HI) combined with the magnetic force theorem and exact diagonalization.  
Since the direct computation of spin exchanges at 
FT-HI and ED for JT-distorted structures is at present not possible, we have first verified the validity of the DFT+U data for the undistorted phase by comparing the spin exchanges extracted from DFT+U total energies  with those calculated explicitly at FT-HI and ED level (Fig.\ref{fig:2}(b)).  
In DFT+U the spin interactions are estimated by fitting the total energies curves obtained at different canting angle (via magnetically constrained calculations) with the spin Hamiltonian given in Eq.~\ref{eq:fitting} (derived in the SM~\cite{supplmat}.
After normalizing the DFT+U data to the FT-HI/ED values (see SM~\cite{supplmat} for further details) we have then addressed the JT-induced modifications at DFT+U level, by performing a series of calculation stepwise from the ideal undistorted phase to the fully distorted real structure
(Fig.\ref{fig:3}(a-b)). In order to account for the role of electron-electron correlation and construct the phase diagram displayed in Fig.~\ref{fig:3}(c), these calculations have been conducted for different values of $U$, from 2.4~eV to 3.4~eV. 

In studying the JT distortion we have followed the Van Vleck notation for the definition of the main JT distortions $Q_1$, $Q_2$ and $Q_3$.
Since the value of $Q_1$ is significantly larger than the corresponding $Q_2$ and $Q_3$ value (see SM~\cite{supplmat}) we have considered the variation of the the spin interactions as a function of the JT parameter $\tilde{Q}_{1}$, defined as the ratio between Q$_{1}$ for the interpolated structure and the corresponding value for the optimal one. 
Additional details can be found in the SM~\cite{supplmat}, which includes Refs.~\cite{PhysRevB.93.214431,PhysRevMaterials.3.014414,Inui1990,PhysRevB.95.205115,PhysRevB.83.220402, PhysRevB.84.115114, leonid,PhysRevB.95.205115,PhysRevB.99.094439}.

\section{Acknowledgements}
We gratefully acknowledge the helpful discussions with V. Mitrovic.
C.F. thanks S. Streltsov for useful insights on the Jahn-Teller effect.
DFM acknowledges support from the Vienna Doctoral School in Physics and from the University of Bologna (Thesis Abroad Scholarship). LVP acknowledges the support by the European Research Council grants ERC-319286-"QMAC" and is grateful to the computer team at CPHT for support. 
B.H.K was supported by a KIAS Individual Grant (CG068701) at Korea Institute for Advanced Study.
The computational results were achieved by using the Vienna Scientific Cluster (VSC). 
\bibliography{biblio}

\begin{thebibliography}{61}%
\makeatletter
\providecommand \@ifxundefined [1]{%
 \@ifx{#1\undefined}
}%
\providecommand \@ifnum [1]{%
 \ifnum #1\expandafter \@firstoftwo
 \else \expandafter \@secondoftwo
 \fi
}%
\providecommand \@ifx [1]{%
 \ifx #1\expandafter \@firstoftwo
 \else \expandafter \@secondoftwo
 \fi
}%
\providecommand \natexlab [1]{#1}%
\providecommand \enquote  [1]{``#1''}%
\providecommand \bibnamefont  [1]{#1}%
\providecommand \bibfnamefont [1]{#1}%
\providecommand \citenamefont [1]{#1}%
\providecommand \href@noop [0]{\@secondoftwo}%
\providecommand \href [0]{\begingroup \@sanitize@url \@href}%
\providecommand \@href[1]{\@@startlink{#1}\@@href}%
\providecommand \@@href[1]{\endgroup#1\@@endlink}%
\providecommand \@sanitize@url [0]{\catcode `\\12\catcode `\$12\catcode
  `\&12\catcode `\#12\catcode `\^12\catcode `\_12\catcode `\%12\relax}%
\providecommand \@@startlink[1]{}%
\providecommand \@@endlink[0]{}%
\providecommand \url  [0]{\begingroup\@sanitize@url \@url }%
\providecommand \@url [1]{\endgroup\@href {#1}{\urlprefix }}%
\providecommand \urlprefix  [0]{URL }%
\providecommand \Eprint [0]{\href }%
\providecommand \doibase [0]{http://dx.doi.org/}%
\providecommand \selectlanguage [0]{\@gobble}%
\providecommand \bibinfo  [0]{\@secondoftwo}%
\providecommand \bibfield  [0]{\@secondoftwo}%
\providecommand \translation [1]{[#1]}%
\providecommand \BibitemOpen [0]{}%
\providecommand \bibitemStop [0]{}%
\providecommand \bibitemNoStop [0]{.\EOS\space}%
\providecommand \EOS [0]{\spacefactor3000\relax}%
\providecommand \BibitemShut  [1]{\csname bibitem#1\endcsname}%
\let\auto@bib@innerbib\@empty
\bibitem [{\citenamefont {Chen}\ \emph {et~al.}(2010)\citenamefont {Chen},
  \citenamefont {Pereira},\ and\ \citenamefont {Balents}}]{PhysRevB.82.174440}%
  \BibitemOpen
  \bibfield  {author} {\bibinfo {author} {\bibfnamefont {G.}~\bibnamefont
  {Chen}}, \bibinfo {author} {\bibfnamefont {R.}~\bibnamefont {Pereira}}, \
  and\ \bibinfo {author} {\bibfnamefont {L.}~\bibnamefont {Balents}},\ }\href
  {\doibase 10.1103/PhysRevB.82.174440} {\bibfield  {journal} {\bibinfo
  {journal} {Phys. Rev. B}\ }\textbf {\bibinfo {volume} {82}},\ \bibinfo
  {pages} {174440} (\bibinfo {year} {2010})}\BibitemShut {NoStop}%
\bibitem [{\citenamefont {Ishizuka}\ and\ \citenamefont
  {Balents}(2014)}]{PhysRevB.90.184422}%
  \BibitemOpen
  \bibfield  {author} {\bibinfo {author} {\bibfnamefont {H.}~\bibnamefont
  {Ishizuka}}\ and\ \bibinfo {author} {\bibfnamefont {L.}~\bibnamefont
  {Balents}},\ }\href {\doibase 10.1103/PhysRevB.90.184422} {\bibfield
  {journal} {\bibinfo  {journal} {Phys. Rev. B}\ }\textbf {\bibinfo {volume}
  {90}},\ \bibinfo {pages} {184422} (\bibinfo {year} {2014})}\BibitemShut
  {NoStop}%
\bibitem [{\citenamefont {Iwahara}\ \emph {et~al.}(2018)\citenamefont
  {Iwahara}, \citenamefont {Vieru},\ and\ \citenamefont
  {Chibotaru}}]{PhysRevB.98.075138}%
  \BibitemOpen
  \bibfield  {author} {\bibinfo {author} {\bibfnamefont {N.}~\bibnamefont
  {Iwahara}}, \bibinfo {author} {\bibfnamefont {V.}~\bibnamefont {Vieru}}, \
  and\ \bibinfo {author} {\bibfnamefont {L.~F.}\ \bibnamefont {Chibotaru}},\
  }\href {\doibase 10.1103/PhysRevB.98.075138} {\bibfield  {journal} {\bibinfo
  {journal} {Phys. Rev. B}\ }\textbf {\bibinfo {volume} {98}},\ \bibinfo
  {pages} {075138} (\bibinfo {year} {2018})}\BibitemShut {NoStop}%
\bibitem [{\citenamefont {Romh\'anyi}\ \emph {et~al.}(2017)\citenamefont
  {Romh\'anyi}, \citenamefont {Balents},\ and\ \citenamefont
  {Jackeli}}]{PhysRevLett.118.217202}%
  \BibitemOpen
  \bibfield  {author} {\bibinfo {author} {\bibfnamefont {J.}~\bibnamefont
  {Romh\'anyi}}, \bibinfo {author} {\bibfnamefont {L.}~\bibnamefont {Balents}},
  \ and\ \bibinfo {author} {\bibfnamefont {G.}~\bibnamefont {Jackeli}},\ }\href
  {\doibase 10.1103/PhysRevLett.118.217202} {\bibfield  {journal} {\bibinfo
  {journal} {Phys. Rev. Lett.}\ }\textbf {\bibinfo {volume} {118}},\ \bibinfo
  {pages} {217202} (\bibinfo {year} {2017})}\BibitemShut {NoStop}%
\bibitem [{\citenamefont {Martins}\ \emph {et~al.}(2017)\citenamefont
  {Martins}, \citenamefont {Aichhorn},\ and\ \citenamefont
  {Biermann}}]{Martins_2017}%
  \BibitemOpen
  \bibfield  {author} {\bibinfo {author} {\bibfnamefont {C.}~\bibnamefont
  {Martins}}, \bibinfo {author} {\bibfnamefont {M.}~\bibnamefont {Aichhorn}}, \
  and\ \bibinfo {author} {\bibfnamefont {S.}~\bibnamefont {Biermann}},\ }\href
  {\doibase 10.1088/1361-648x/aa648f} {\bibfield  {journal} {\bibinfo
  {journal} {Journal of Physics: Condensed Matter}\ }\textbf {\bibinfo {volume}
  {29}},\ \bibinfo {pages} {263001} (\bibinfo {year} {2017})}\BibitemShut
  {NoStop}%
\bibitem [{\citenamefont {Khomskii}\ and\ \citenamefont
  {Mostovoy}(2003)}]{Khomskii_2003}%
  \BibitemOpen
  \bibfield  {author} {\bibinfo {author} {\bibfnamefont {D.~I.}\ \bibnamefont
  {Khomskii}}\ and\ \bibinfo {author} {\bibfnamefont {M.~V.}\ \bibnamefont
  {Mostovoy}},\ }\href {\doibase 10.1088/0305-4470/36/35/307} {\bibfield
  {journal} {\bibinfo  {journal} {Journal of Physics A: Mathematical and
  General}\ }\textbf {\bibinfo {volume} {36}},\ \bibinfo {pages} {9197}
  (\bibinfo {year} {2003})}\BibitemShut {NoStop}%
\bibitem [{\citenamefont {Goodenough}(1968)}]{PhysRev.171.466}%
  \BibitemOpen
  \bibfield  {author} {\bibinfo {author} {\bibfnamefont {J.~B.}\ \bibnamefont
  {Goodenough}},\ }\href {\doibase 10.1103/PhysRev.171.466} {\bibfield
  {journal} {\bibinfo  {journal} {Phys. Rev.}\ }\textbf {\bibinfo {volume}
  {171}},\ \bibinfo {pages} {466} (\bibinfo {year} {1968})}\BibitemShut
  {NoStop}%
\bibitem [{\citenamefont {Lee}\ and\ \citenamefont {Pickett}(2007)}]{Lee_2007}%
  \BibitemOpen
  \bibfield  {author} {\bibinfo {author} {\bibfnamefont {K.-W.}\ \bibnamefont
  {Lee}}\ and\ \bibinfo {author} {\bibfnamefont {W.~E.}\ \bibnamefont
  {Pickett}},\ }\href {\doibase 10.1209/0295-5075/80/37008} {\bibfield
  {journal} {\bibinfo  {journal} {Europhysics Letters ({EPL})}\ }\textbf
  {\bibinfo {volume} {80}},\ \bibinfo {pages} {37008} (\bibinfo {year}
  {2007})}\BibitemShut {NoStop}%
\bibitem [{\citenamefont {Xiang}\ and\ \citenamefont
  {Whangbo}(2007)}]{PhysRevB.75.052407}%
  \BibitemOpen
  \bibfield  {author} {\bibinfo {author} {\bibfnamefont {H.~J.}\ \bibnamefont
  {Xiang}}\ and\ \bibinfo {author} {\bibfnamefont {M.-H.}\ \bibnamefont
  {Whangbo}},\ }\href {\doibase 10.1103/PhysRevB.75.052407} {\bibfield
  {journal} {\bibinfo  {journal} {Phys. Rev. B}\ }\textbf {\bibinfo {volume}
  {75}},\ \bibinfo {pages} {052407} (\bibinfo {year} {2007})}\BibitemShut
  {NoStop}%
\bibitem [{\citenamefont {Xu}\ \emph {et~al.}(2016)\citenamefont {Xu},
  \citenamefont {Bogdanov}, \citenamefont {Princep}, \citenamefont {Fulde},
  \citenamefont {van~den Brink},\ and\ \citenamefont {Hozoi}}]{Xu2016}%
  \BibitemOpen
  \bibfield  {author} {\bibinfo {author} {\bibfnamefont {L.}~\bibnamefont
  {Xu}}, \bibinfo {author} {\bibfnamefont {N.~A.}\ \bibnamefont {Bogdanov}},
  \bibinfo {author} {\bibfnamefont {A.}~\bibnamefont {Princep}}, \bibinfo
  {author} {\bibfnamefont {P.}~\bibnamefont {Fulde}}, \bibinfo {author}
  {\bibfnamefont {J.}~\bibnamefont {van~den Brink}}, \ and\ \bibinfo {author}
  {\bibfnamefont {L.}~\bibnamefont {Hozoi}},\ }\href {\doibase
  10.1038/npjquantmats.2016.29} {\bibfield  {journal} {\bibinfo  {journal} {npj
  Quantum Materials}\ }\textbf {\bibinfo {volume} {1}},\ \bibinfo {pages}
  {16029} (\bibinfo {year} {2016})}\BibitemShut {NoStop}%
\bibitem [{\citenamefont {Feng}\ \emph {et~al.}(2016)\citenamefont {Feng},
  \citenamefont {Calder}, \citenamefont {Ghimire}, \citenamefont {Yuan},
  \citenamefont {Shirako}, \citenamefont {Tsujimoto}, \citenamefont
  {Matsushita}, \citenamefont {Hu}, \citenamefont {Kuo}, \citenamefont {Tjeng},
  \citenamefont {Pi}, \citenamefont {Soo}, \citenamefont {He}, \citenamefont
  {Tanaka}, \citenamefont {Katsuya}, \citenamefont {Richter},\ and\
  \citenamefont {Yamaura}}]{PhysRevB.94.235158}%
  \BibitemOpen
  \bibfield  {author} {\bibinfo {author} {\bibfnamefont {H.~L.}\ \bibnamefont
  {Feng}}, \bibinfo {author} {\bibfnamefont {S.}~\bibnamefont {Calder}},
  \bibinfo {author} {\bibfnamefont {M.~P.}\ \bibnamefont {Ghimire}}, \bibinfo
  {author} {\bibfnamefont {Y.-H.}\ \bibnamefont {Yuan}}, \bibinfo {author}
  {\bibfnamefont {Y.}~\bibnamefont {Shirako}}, \bibinfo {author} {\bibfnamefont
  {Y.}~\bibnamefont {Tsujimoto}}, \bibinfo {author} {\bibfnamefont
  {Y.}~\bibnamefont {Matsushita}}, \bibinfo {author} {\bibfnamefont
  {Z.}~\bibnamefont {Hu}}, \bibinfo {author} {\bibfnamefont {C.-Y.}\
  \bibnamefont {Kuo}}, \bibinfo {author} {\bibfnamefont {L.~H.}\ \bibnamefont
  {Tjeng}}, \bibinfo {author} {\bibfnamefont {T.-W.}\ \bibnamefont {Pi}},
  \bibinfo {author} {\bibfnamefont {Y.-L.}\ \bibnamefont {Soo}}, \bibinfo
  {author} {\bibfnamefont {J.}~\bibnamefont {He}}, \bibinfo {author}
  {\bibfnamefont {M.}~\bibnamefont {Tanaka}}, \bibinfo {author} {\bibfnamefont
  {Y.}~\bibnamefont {Katsuya}}, \bibinfo {author} {\bibfnamefont
  {M.}~\bibnamefont {Richter}}, \ and\ \bibinfo {author} {\bibfnamefont
  {K.}~\bibnamefont {Yamaura}},\ }\href {\doibase 10.1103/PhysRevB.94.235158}
  {\bibfield  {journal} {\bibinfo  {journal} {Phys. Rev. B}\ }\textbf {\bibinfo
  {volume} {94}},\ \bibinfo {pages} {235158} (\bibinfo {year}
  {2016})}\BibitemShut {NoStop}%
\bibitem [{\citenamefont {Gangopadhyay}\ and\ \citenamefont
  {Pickett}(2015)}]{PhysRevB.91.045133}%
  \BibitemOpen
  \bibfield  {author} {\bibinfo {author} {\bibfnamefont {S.}~\bibnamefont
  {Gangopadhyay}}\ and\ \bibinfo {author} {\bibfnamefont {W.~E.}\ \bibnamefont
  {Pickett}},\ }\href {\doibase 10.1103/PhysRevB.91.045133} {\bibfield
  {journal} {\bibinfo  {journal} {Phys. Rev. B}\ }\textbf {\bibinfo {volume}
  {91}},\ \bibinfo {pages} {045133} (\bibinfo {year} {2015})}\BibitemShut
  {NoStop}%
\bibitem [{\citenamefont {Gangopadhyay}\ and\ \citenamefont
  {Pickett}(2016)}]{PhysRevB.93.155126}%
  \BibitemOpen
  \bibfield  {author} {\bibinfo {author} {\bibfnamefont {S.}~\bibnamefont
  {Gangopadhyay}}\ and\ \bibinfo {author} {\bibfnamefont {W.~E.}\ \bibnamefont
  {Pickett}},\ }\href {\doibase 10.1103/PhysRevB.93.155126} {\bibfield
  {journal} {\bibinfo  {journal} {Phys. Rev. B}\ }\textbf {\bibinfo {volume}
  {93}},\ \bibinfo {pages} {155126} (\bibinfo {year} {2016})}\BibitemShut
  {NoStop}%
\bibitem [{\citenamefont {Wang}\ \emph {et~al.}(2019)\citenamefont {Wang},
  \citenamefont {Xu}, \citenamefont {Hao}, \citenamefont {Wang}, \citenamefont
  {Zhang}, \citenamefont {Sun},\ and\ \citenamefont
  {Hao}}]{PhysRevB.99.035126}%
  \BibitemOpen
  \bibfield  {author} {\bibinfo {author} {\bibfnamefont {C.}~\bibnamefont
  {Wang}}, \bibinfo {author} {\bibfnamefont {Y.}~\bibnamefont {Xu}}, \bibinfo
  {author} {\bibfnamefont {W.}~\bibnamefont {Hao}}, \bibinfo {author}
  {\bibfnamefont {R.}~\bibnamefont {Wang}}, \bibinfo {author} {\bibfnamefont
  {X.}~\bibnamefont {Zhang}}, \bibinfo {author} {\bibfnamefont
  {K.}~\bibnamefont {Sun}}, \ and\ \bibinfo {author} {\bibfnamefont
  {X.}~\bibnamefont {Hao}},\ }\href {\doibase 10.1103/PhysRevB.99.035126}
  {\bibfield  {journal} {\bibinfo  {journal} {Phys. Rev. B}\ }\textbf {\bibinfo
  {volume} {99}},\ \bibinfo {pages} {035126} (\bibinfo {year}
  {2019})}\BibitemShut {NoStop}%
\bibitem [{\citenamefont {Stitzer}\ \emph {et~al.}(2002)\citenamefont
  {Stitzer}, \citenamefont {Smith},\ and\ \citenamefont {zur
  Loye}}]{STITZER2002311}%
  \BibitemOpen
  \bibfield  {author} {\bibinfo {author} {\bibfnamefont {K.~E.}\ \bibnamefont
  {Stitzer}}, \bibinfo {author} {\bibfnamefont {M.~D.}\ \bibnamefont {Smith}},
  \ and\ \bibinfo {author} {\bibfnamefont {H.-C.}\ \bibnamefont {zur Loye}},\
  }\href {\doibase https://doi.org/10.1016/S1293-2558(01)01257-2} {\bibfield
  {journal} {\bibinfo  {journal} {Solid State Sciences}\ }\textbf {\bibinfo
  {volume} {4}},\ \bibinfo {pages} {311 } (\bibinfo {year} {2002})}\BibitemShut
  {NoStop}%
\bibitem [{\citenamefont {Erickson}\ \emph {et~al.}(2007)\citenamefont
  {Erickson}, \citenamefont {Misra}, \citenamefont {Miller}, \citenamefont
  {Gupta}, \citenamefont {Schlesinger}, \citenamefont {Harrison}, \citenamefont
  {Kim},\ and\ \citenamefont {Fisher}}]{PhysRevLett.99.016404}%
  \BibitemOpen
  \bibfield  {author} {\bibinfo {author} {\bibfnamefont {A.~S.}\ \bibnamefont
  {Erickson}}, \bibinfo {author} {\bibfnamefont {S.}~\bibnamefont {Misra}},
  \bibinfo {author} {\bibfnamefont {G.~J.}\ \bibnamefont {Miller}}, \bibinfo
  {author} {\bibfnamefont {R.~R.}\ \bibnamefont {Gupta}}, \bibinfo {author}
  {\bibfnamefont {Z.}~\bibnamefont {Schlesinger}}, \bibinfo {author}
  {\bibfnamefont {W.~A.}\ \bibnamefont {Harrison}}, \bibinfo {author}
  {\bibfnamefont {J.~M.}\ \bibnamefont {Kim}}, \ and\ \bibinfo {author}
  {\bibfnamefont {I.~R.}\ \bibnamefont {Fisher}},\ }\href {\doibase
  10.1103/PhysRevLett.99.016404} {\bibfield  {journal} {\bibinfo  {journal}
  {Phys. Rev. Lett.}\ }\textbf {\bibinfo {volume} {99}},\ \bibinfo {pages}
  {016404} (\bibinfo {year} {2007})}\BibitemShut {NoStop}%
\bibitem [{\citenamefont {Steele}\ \emph {et~al.}(2011)\citenamefont {Steele},
  \citenamefont {Baker}, \citenamefont {Lancaster}, \citenamefont {Pratt},
  \citenamefont {Franke}, \citenamefont {Ghannadzadeh}, \citenamefont
  {Goddard}, \citenamefont {Hayes}, \citenamefont {Prabhakaran},\ and\
  \citenamefont {Blundell}}]{PhysRevB.84.144416}%
  \BibitemOpen
  \bibfield  {author} {\bibinfo {author} {\bibfnamefont {A.~J.}\ \bibnamefont
  {Steele}}, \bibinfo {author} {\bibfnamefont {P.~J.}\ \bibnamefont {Baker}},
  \bibinfo {author} {\bibfnamefont {T.}~\bibnamefont {Lancaster}}, \bibinfo
  {author} {\bibfnamefont {F.~L.}\ \bibnamefont {Pratt}}, \bibinfo {author}
  {\bibfnamefont {I.}~\bibnamefont {Franke}}, \bibinfo {author} {\bibfnamefont
  {S.}~\bibnamefont {Ghannadzadeh}}, \bibinfo {author} {\bibfnamefont {P.~A.}\
  \bibnamefont {Goddard}}, \bibinfo {author} {\bibfnamefont {W.}~\bibnamefont
  {Hayes}}, \bibinfo {author} {\bibfnamefont {D.}~\bibnamefont {Prabhakaran}},
  \ and\ \bibinfo {author} {\bibfnamefont {S.~J.}\ \bibnamefont {Blundell}},\
  }\href {\doibase 10.1103/PhysRevB.84.144416} {\bibfield  {journal} {\bibinfo
  {journal} {Phys. Rev. B}\ }\textbf {\bibinfo {volume} {84}},\ \bibinfo
  {pages} {144416} (\bibinfo {year} {2011})}\BibitemShut {NoStop}%
\bibitem [{\citenamefont {Ahn}\ \emph {et~al.}(2017)\citenamefont {Ahn},
  \citenamefont {Pajskr}, \citenamefont {Lee},\ and\ \citenamefont
  {Kune\ifmmode~\check{s}\else \v{s}\fi{}}}]{PhysRevB.95.064416}%
  \BibitemOpen
  \bibfield  {author} {\bibinfo {author} {\bibfnamefont {K.-H.}\ \bibnamefont
  {Ahn}}, \bibinfo {author} {\bibfnamefont {K.}~\bibnamefont {Pajskr}},
  \bibinfo {author} {\bibfnamefont {K.-W.}\ \bibnamefont {Lee}}, \ and\
  \bibinfo {author} {\bibfnamefont {J.}~\bibnamefont
  {Kune\ifmmode~\check{s}\else \v{s}\fi{}}},\ }\href {\doibase
  10.1103/PhysRevB.95.064416} {\bibfield  {journal} {\bibinfo  {journal} {Phys.
  Rev. B}\ }\textbf {\bibinfo {volume} {95}},\ \bibinfo {pages} {064416}
  (\bibinfo {year} {2017})}\BibitemShut {NoStop}%
\bibitem [{\citenamefont {Lu}\ \emph {et~al.}(2017)\citenamefont {Lu},
  \citenamefont {Song}, \citenamefont {Liu}, \citenamefont {Reyes},
  \citenamefont {Kuhns}, \citenamefont {Lee}, \citenamefont {Fisher},\ and\
  \citenamefont {Mitrovic}}]{Lu2017}%
  \BibitemOpen
  \bibfield  {author} {\bibinfo {author} {\bibfnamefont {L.}~\bibnamefont
  {Lu}}, \bibinfo {author} {\bibfnamefont {M.}~\bibnamefont {Song}}, \bibinfo
  {author} {\bibfnamefont {W.}~\bibnamefont {Liu}}, \bibinfo {author}
  {\bibfnamefont {A.~P.}\ \bibnamefont {Reyes}}, \bibinfo {author}
  {\bibfnamefont {P.}~\bibnamefont {Kuhns}}, \bibinfo {author} {\bibfnamefont
  {H.~O.}\ \bibnamefont {Lee}}, \bibinfo {author} {\bibfnamefont {I.~R.}\
  \bibnamefont {Fisher}}, \ and\ \bibinfo {author} {\bibfnamefont {V.~F.}\
  \bibnamefont {Mitrovic}},\ }\href {\doibase 10.1038/ncomms14407} {\bibfield
  {journal} {\bibinfo  {journal} {Nature Communications}\ }\textbf {\bibinfo
  {volume} {8}},\ \bibinfo {pages} {14407} (\bibinfo {year}
  {2017})}\BibitemShut {NoStop}%
\bibitem [{\citenamefont {Willa}\ \emph {et~al.}(2019)\citenamefont {Willa},
  \citenamefont {Willa}, \citenamefont {Welp}, \citenamefont {Fisher},
  \citenamefont {Rydh}, \citenamefont {Kwok},\ and\ \citenamefont
  {Islam}}]{PhysRevB.100.041108}%
  \BibitemOpen
  \bibfield  {author} {\bibinfo {author} {\bibfnamefont {K.}~\bibnamefont
  {Willa}}, \bibinfo {author} {\bibfnamefont {R.}~\bibnamefont {Willa}},
  \bibinfo {author} {\bibfnamefont {U.}~\bibnamefont {Welp}}, \bibinfo {author}
  {\bibfnamefont {I.~R.}\ \bibnamefont {Fisher}}, \bibinfo {author}
  {\bibfnamefont {A.}~\bibnamefont {Rydh}}, \bibinfo {author} {\bibfnamefont
  {W.-K.}\ \bibnamefont {Kwok}}, \ and\ \bibinfo {author} {\bibfnamefont
  {Z.}~\bibnamefont {Islam}},\ }\href {\doibase 10.1103/PhysRevB.100.041108}
  {\bibfield  {journal} {\bibinfo  {journal} {Phys. Rev. B}\ }\textbf {\bibinfo
  {volume} {100}},\ \bibinfo {pages} {041108} (\bibinfo {year}
  {2019})}\BibitemShut {NoStop}%
\bibitem [{\citenamefont {Liu}\ \emph {et~al.}(2018)\citenamefont {Liu},
  \citenamefont {Cong}, \citenamefont {Reyes}, \citenamefont {Fisher},\ and\
  \citenamefont {Mitrovi\ifmmode~\acute{c}\else
  \'{c}\fi{}}}]{PhysRevB.97.224103}%
  \BibitemOpen
  \bibfield  {author} {\bibinfo {author} {\bibfnamefont {W.}~\bibnamefont
  {Liu}}, \bibinfo {author} {\bibfnamefont {R.}~\bibnamefont {Cong}}, \bibinfo
  {author} {\bibfnamefont {A.~P.}\ \bibnamefont {Reyes}}, \bibinfo {author}
  {\bibfnamefont {I.~R.}\ \bibnamefont {Fisher}}, \ and\ \bibinfo {author}
  {\bibfnamefont {V.~F.}\ \bibnamefont {Mitrovi\ifmmode~\acute{c}\else
  \'{c}\fi{}}},\ }\href {\doibase 10.1103/PhysRevB.97.224103} {\bibfield
  {journal} {\bibinfo  {journal} {Phys. Rev. B}\ }\textbf {\bibinfo {volume}
  {97}},\ \bibinfo {pages} {224103} (\bibinfo {year} {2018})}\BibitemShut
  {NoStop}%
\bibitem [{\citenamefont {Khomskii}(2014)}]{khombook}%
  \BibitemOpen
  \bibfield  {author} {\bibinfo {author} {\bibfnamefont {D.~I.}\ \bibnamefont
  {Khomskii}},\ }\href {\doibase 10.1017/CBO9781139096782} {\emph {\bibinfo
  {title} {Transition Metal Compounds}}}\ (\bibinfo  {publisher} {Cambridge
  University Press},\ \bibinfo {year} {2014})\BibitemShut {NoStop}%
\bibitem [{\citenamefont {Jahn}\ and\ \citenamefont {Bragg}(1938)}]{jt2}%
  \BibitemOpen
  \bibfield  {author} {\bibinfo {author} {\bibfnamefont {H.~A.}\ \bibnamefont
  {Jahn}}\ and\ \bibinfo {author} {\bibfnamefont {W.~H.}\ \bibnamefont
  {Bragg}},\ }\href {\doibase 10.1098/rspa.1938.0008} {\bibfield  {journal}
  {\bibinfo  {journal} {Proceedings of the Royal Society of London. Series A -
  Mathematical and Physical Sciences}\ }\textbf {\bibinfo {volume} {164}},\
  \bibinfo {pages} {117} (\bibinfo {year} {1938})}\BibitemShut {NoStop}%
\bibitem [{\citenamefont {Gehring}\ and\ \citenamefont
  {Gehring}(1975)}]{Gehring_1975}%
  \BibitemOpen
  \bibfield  {author} {\bibinfo {author} {\bibfnamefont {G.~A.}\ \bibnamefont
  {Gehring}}\ and\ \bibinfo {author} {\bibfnamefont {K.~A.}\ \bibnamefont
  {Gehring}},\ }\href {\doibase 10.1088/0034-4885/38/1/001} {\bibfield
  {journal} {\bibinfo  {journal} {Reports on Progress in Physics}\ }\textbf
  {\bibinfo {volume} {38}},\ \bibinfo {pages} {1} (\bibinfo {year}
  {1975})}\BibitemShut {NoStop}%
\bibitem [{\citenamefont {Cong}\ \emph {et~al.}(2019)\citenamefont {Cong},
  \citenamefont {Nanguneri}, \citenamefont {Rubenstein},\ and\ \citenamefont
  {Mitrovi\ifmmode~\acute{c}\else \'{c}\fi{}}}]{PhysRevB.100.245141}%
  \BibitemOpen
  \bibfield  {author} {\bibinfo {author} {\bibfnamefont {R.}~\bibnamefont
  {Cong}}, \bibinfo {author} {\bibfnamefont {R.}~\bibnamefont {Nanguneri}},
  \bibinfo {author} {\bibfnamefont {B.}~\bibnamefont {Rubenstein}}, \ and\
  \bibinfo {author} {\bibfnamefont {V.~F.}\ \bibnamefont
  {Mitrovi\ifmmode~\acute{c}\else \'{c}\fi{}}},\ }\href {\doibase
  10.1103/PhysRevB.100.245141} {\bibfield  {journal} {\bibinfo  {journal}
  {Phys. Rev. B}\ }\textbf {\bibinfo {volume} {100}},\ \bibinfo {pages}
  {245141} (\bibinfo {year} {2019})}\BibitemShut {NoStop}%
\bibitem [{\citenamefont {Van~Vleck}(1939)}]{Vleck1939}%
  \BibitemOpen
  \bibfield  {author} {\bibinfo {author} {\bibfnamefont {J.~H.}\ \bibnamefont
  {Van~Vleck}},\ }\href {\doibase 10.1063/1.1750327} {\bibfield  {journal}
  {\bibinfo  {journal} {The Journal of Chemical Physics}\ }\textbf {\bibinfo
  {volume} {7}},\ \bibinfo {pages} {72} (\bibinfo {year} {1939})},\ \Eprint
  {http://arxiv.org/abs/https://doi.org/10.1063/1.1750327}
  {https://doi.org/10.1063/1.1750327} \BibitemShut {NoStop}%
\bibitem [{sup()}]{supplmat}%
  \BibitemOpen
  \href@noop {} {}\bibinfo {note} {Supplementary Material}\BibitemShut
  {NoStop}%
\bibitem [{\citenamefont {Kanamori}(1960)}]{doi:10.1063/1.1984590}%
  \BibitemOpen
  \bibfield  {author} {\bibinfo {author} {\bibfnamefont {J.}~\bibnamefont
  {Kanamori}},\ }\href {\doibase 10.1063/1.1984590} {\bibfield  {journal}
  {\bibinfo  {journal} {Journal of Applied Physics}\ }\textbf {\bibinfo
  {volume} {31}},\ \bibinfo {pages} {S14} (\bibinfo {year} {1960})},\ \Eprint
  {http://arxiv.org/abs/https://doi.org/10.1063/1.1984590}
  {https://doi.org/10.1063/1.1984590} \BibitemShut {NoStop}%
\bibitem [{\citenamefont {Hobbs}\ \emph {et~al.}(2000)\citenamefont {Hobbs},
  \citenamefont {Kresse},\ and\ \citenamefont {Hafner}}]{PhysRevB.62.11556}%
  \BibitemOpen
  \bibfield  {author} {\bibinfo {author} {\bibfnamefont {D.}~\bibnamefont
  {Hobbs}}, \bibinfo {author} {\bibfnamefont {G.}~\bibnamefont {Kresse}}, \
  and\ \bibinfo {author} {\bibfnamefont {J.}~\bibnamefont {Hafner}},\ }\href
  {\doibase 10.1103/PhysRevB.62.11556} {\bibfield  {journal} {\bibinfo
  {journal} {Phys. Rev. B}\ }\textbf {\bibinfo {volume} {62}},\ \bibinfo
  {pages} {11556} (\bibinfo {year} {2000})}\BibitemShut {NoStop}%
\bibitem [{\citenamefont {Liu}\ \emph {et~al.}(2015)\citenamefont {Liu},
  \citenamefont {Khmelevskyi}, \citenamefont {Kim}, \citenamefont {Marsman},
  \citenamefont {Li}, \citenamefont {Chen}, \citenamefont {Sarma},
  \citenamefont {Kresse},\ and\ \citenamefont {Franchini}}]{Liu2015}%
  \BibitemOpen
  \bibfield  {author} {\bibinfo {author} {\bibfnamefont {P.}~\bibnamefont
  {Liu}}, \bibinfo {author} {\bibfnamefont {S.}~\bibnamefont {Khmelevskyi}},
  \bibinfo {author} {\bibfnamefont {B.}~\bibnamefont {Kim}}, \bibinfo {author}
  {\bibfnamefont {M.}~\bibnamefont {Marsman}}, \bibinfo {author} {\bibfnamefont
  {D.}~\bibnamefont {Li}}, \bibinfo {author} {\bibfnamefont {X.-Q.}\
  \bibnamefont {Chen}}, \bibinfo {author} {\bibfnamefont {D.~D.}\ \bibnamefont
  {Sarma}}, \bibinfo {author} {\bibfnamefont {G.}~\bibnamefont {Kresse}}, \
  and\ \bibinfo {author} {\bibfnamefont {C.}~\bibnamefont {Franchini}},\ }\href
  {\doibase 10.1103/PhysRevB.92.054428} {\bibfield  {journal} {\bibinfo
  {journal} {Phys. Rev. B}\ }\textbf {\bibinfo {volume} {92}},\ \bibinfo
  {pages} {054428} (\bibinfo {year} {2015})}\BibitemShut {NoStop}%
\bibitem [{\citenamefont {Dudarev}\ \emph {et~al.}(2019)\citenamefont
  {Dudarev}, \citenamefont {Liu}, \citenamefont {Andersson}, \citenamefont
  {Stanek}, \citenamefont {Ozaki},\ and\ \citenamefont
  {Franchini}}]{PhysRevMaterials.3.083802}%
  \BibitemOpen
  \bibfield  {author} {\bibinfo {author} {\bibfnamefont {S.~L.}\ \bibnamefont
  {Dudarev}}, \bibinfo {author} {\bibfnamefont {P.}~\bibnamefont {Liu}},
  \bibinfo {author} {\bibfnamefont {D.~A.}\ \bibnamefont {Andersson}}, \bibinfo
  {author} {\bibfnamefont {C.~R.}\ \bibnamefont {Stanek}}, \bibinfo {author}
  {\bibfnamefont {T.}~\bibnamefont {Ozaki}}, \ and\ \bibinfo {author}
  {\bibfnamefont {C.}~\bibnamefont {Franchini}},\ }\href {\doibase
  10.1103/PhysRevMaterials.3.083802} {\bibfield  {journal} {\bibinfo  {journal}
  {Phys. Rev. Materials}\ }\textbf {\bibinfo {volume} {3}},\ \bibinfo {pages}
  {083802} (\bibinfo {year} {2019})}\BibitemShut {NoStop}%
\bibitem [{\citenamefont {Perdew}\ \emph {et~al.}(1996)\citenamefont {Perdew},
  \citenamefont {Burke},\ and\ \citenamefont
  {Ernzerhof}}]{PhysRevLett.77.3865}%
  \BibitemOpen
  \bibfield  {author} {\bibinfo {author} {\bibfnamefont {J.~P.}\ \bibnamefont
  {Perdew}}, \bibinfo {author} {\bibfnamefont {K.}~\bibnamefont {Burke}}, \
  and\ \bibinfo {author} {\bibfnamefont {M.}~\bibnamefont {Ernzerhof}},\ }\href
  {\doibase 10.1103/PhysRevLett.77.3865} {\bibfield  {journal} {\bibinfo
  {journal} {Phys. Rev. Lett.}\ }\textbf {\bibinfo {volume} {77}},\ \bibinfo
  {pages} {3865} (\bibinfo {year} {1996})}\BibitemShut {NoStop}%
\bibitem [{\citenamefont {Kresse}\ and\ \citenamefont
  {Hafner}(1993)}]{PhysRevB.47.558}%
  \BibitemOpen
  \bibfield  {author} {\bibinfo {author} {\bibfnamefont {G.}~\bibnamefont
  {Kresse}}\ and\ \bibinfo {author} {\bibfnamefont {J.}~\bibnamefont
  {Hafner}},\ }\href {\doibase 10.1103/PhysRevB.47.558} {\bibfield  {journal}
  {\bibinfo  {journal} {Phys. Rev. B}\ }\textbf {\bibinfo {volume} {47}},\
  \bibinfo {pages} {558} (\bibinfo {year} {1993})}\BibitemShut {NoStop}%
\bibitem [{\citenamefont {Kresse}\ and\ \citenamefont
  {Furthm\"uller}(1996)}]{PhysRevB.54.11169}%
  \BibitemOpen
  \bibfield  {author} {\bibinfo {author} {\bibfnamefont {G.}~\bibnamefont
  {Kresse}}\ and\ \bibinfo {author} {\bibfnamefont {J.}~\bibnamefont
  {Furthm\"uller}},\ }\href {\doibase 10.1103/PhysRevB.54.11169} {\bibfield
  {journal} {\bibinfo  {journal} {Phys. Rev. B}\ }\textbf {\bibinfo {volume}
  {54}},\ \bibinfo {pages} {11169} (\bibinfo {year} {1996})}\BibitemShut
  {NoStop}%
\bibitem [{\citenamefont {Bl\"ochl}(1994)}]{PhysRevB.50.17953}%
  \BibitemOpen
  \bibfield  {author} {\bibinfo {author} {\bibfnamefont {P.~E.}\ \bibnamefont
  {Bl\"ochl}},\ }\href {\doibase 10.1103/PhysRevB.50.17953} {\bibfield
  {journal} {\bibinfo  {journal} {Phys. Rev. B}\ }\textbf {\bibinfo {volume}
  {50}},\ \bibinfo {pages} {17953} (\bibinfo {year} {1994})}\BibitemShut
  {NoStop}%
\bibitem [{\citenamefont {Georges}\ \emph {et~al.}(1996)\citenamefont
  {Georges}, \citenamefont {Kotliar}, \citenamefont {Krauth},\ and\
  \citenamefont {Rozenberg}}]{RevModPhys.68.13}%
  \BibitemOpen
  \bibfield  {author} {\bibinfo {author} {\bibfnamefont {A.}~\bibnamefont
  {Georges}}, \bibinfo {author} {\bibfnamefont {G.}~\bibnamefont {Kotliar}},
  \bibinfo {author} {\bibfnamefont {W.}~\bibnamefont {Krauth}}, \ and\ \bibinfo
  {author} {\bibfnamefont {M.~J.}\ \bibnamefont {Rozenberg}},\ }\href {\doibase
  10.1103/RevModPhys.68.13} {\bibfield  {journal} {\bibinfo  {journal} {Rev.
  Mod. Phys.}\ }\textbf {\bibinfo {volume} {68}},\ \bibinfo {pages} {13}
  (\bibinfo {year} {1996})}\BibitemShut {NoStop}%
\bibitem [{\citenamefont {Anisimov}\ \emph {et~al.}(1997)\citenamefont
  {Anisimov}, \citenamefont {Poteryaev}, \citenamefont {Korotin}, \citenamefont
  {Anokhin},\ and\ \citenamefont {Kotliar}}]{Anisimov_1997}%
  \BibitemOpen
  \bibfield  {author} {\bibinfo {author} {\bibfnamefont {V.~I.}\ \bibnamefont
  {Anisimov}}, \bibinfo {author} {\bibfnamefont {A.~I.}\ \bibnamefont
  {Poteryaev}}, \bibinfo {author} {\bibfnamefont {M.~A.}\ \bibnamefont
  {Korotin}}, \bibinfo {author} {\bibfnamefont {A.~O.}\ \bibnamefont
  {Anokhin}}, \ and\ \bibinfo {author} {\bibfnamefont {G.}~\bibnamefont
  {Kotliar}},\ }\href {\doibase 10.1088/0953-8984/9/35/010} {\bibfield
  {journal} {\bibinfo  {journal} {Journal of Physics: Condensed Matter}\
  }\textbf {\bibinfo {volume} {9}},\ \bibinfo {pages} {7359} (\bibinfo {year}
  {1997})}\BibitemShut {NoStop}%
\bibitem [{\citenamefont {Hubbard}(1963)}]{hubbard_1}%
  \BibitemOpen
  \bibfield  {author} {\bibinfo {author} {\bibfnamefont {J.}~\bibnamefont
  {Hubbard}},\ }\href@noop {} {\bibfield  {journal} {\bibinfo  {journal} {Proc.
  Roy. Soc. (London)}\ }\textbf {\bibinfo {volume} {A 276}},\ \bibinfo {pages}
  {238} (\bibinfo {year} {1963})}\BibitemShut {NoStop}%
\bibitem [{\citenamefont {Lichtenstein}\ and\ \citenamefont
  {Katsnelson}(1998)}]{lichtenstein_dft_dmft}%
  \BibitemOpen
  \bibfield  {author} {\bibinfo {author} {\bibfnamefont {A.~I.}\ \bibnamefont
  {Lichtenstein}}\ and\ \bibinfo {author} {\bibfnamefont {M.~I.}\ \bibnamefont
  {Katsnelson}},\ }\href@noop {} {\bibfield  {journal} {\bibinfo  {journal}
  {Phys. Rev. B}\ }\textbf {\bibinfo {volume} {57}},\ \bibinfo {pages} {6884}
  (\bibinfo {year} {1998})}\BibitemShut {NoStop}%
\bibitem [{\citenamefont {Pourovskii}(2016)}]{PhysRevB.94.115117}%
  \BibitemOpen
  \bibfield  {author} {\bibinfo {author} {\bibfnamefont {L.~V.}\ \bibnamefont
  {Pourovskii}},\ }\href {\doibase 10.1103/PhysRevB.94.115117} {\bibfield
  {journal} {\bibinfo  {journal} {Phys. Rev. B}\ }\textbf {\bibinfo {volume}
  {94}},\ \bibinfo {pages} {115117} (\bibinfo {year} {2016})}\BibitemShut
  {NoStop}%
\bibitem [{\citenamefont {Kim}\ \emph {et~al.}(2012)\citenamefont {Kim},
  \citenamefont {Khaliullin},\ and\ \citenamefont
  {Min}}]{PhysRevLett.109.167205}%
  \BibitemOpen
  \bibfield  {author} {\bibinfo {author} {\bibfnamefont {B.~H.}\ \bibnamefont
  {Kim}}, \bibinfo {author} {\bibfnamefont {G.}~\bibnamefont {Khaliullin}}, \
  and\ \bibinfo {author} {\bibfnamefont {B.~I.}\ \bibnamefont {Min}},\ }\href
  {\doibase 10.1103/PhysRevLett.109.167205} {\bibfield  {journal} {\bibinfo
  {journal} {Phys. Rev. Lett.}\ }\textbf {\bibinfo {volume} {109}},\ \bibinfo
  {pages} {167205} (\bibinfo {year} {2012})}\BibitemShut {NoStop}%
\bibitem [{\citenamefont {Aichhorn}\ \emph {et~al.}(2009)\citenamefont
  {Aichhorn}, \citenamefont {Pourovskii}, \citenamefont {Vildosola},
  \citenamefont {Ferrero}, \citenamefont {Parcollet}, \citenamefont {Miyake},
  \citenamefont {Georges},\ and\ \citenamefont {Biermann}}]{Aichhorn2009}%
  \BibitemOpen
  \bibfield  {author} {\bibinfo {author} {\bibfnamefont {M.}~\bibnamefont
  {Aichhorn}}, \bibinfo {author} {\bibfnamefont {L.}~\bibnamefont
  {Pourovskii}}, \bibinfo {author} {\bibfnamefont {V.}~\bibnamefont
  {Vildosola}}, \bibinfo {author} {\bibfnamefont {M.}~\bibnamefont {Ferrero}},
  \bibinfo {author} {\bibfnamefont {O.}~\bibnamefont {Parcollet}}, \bibinfo
  {author} {\bibfnamefont {T.}~\bibnamefont {Miyake}}, \bibinfo {author}
  {\bibfnamefont {A.}~\bibnamefont {Georges}}, \ and\ \bibinfo {author}
  {\bibfnamefont {S.}~\bibnamefont {Biermann}},\ }\href@noop {} {\bibfield
  {journal} {\bibinfo  {journal} {Phys. Rev. B}\ }\textbf {\bibinfo {volume}
  {80}},\ \bibinfo {pages} {085101} (\bibinfo {year} {2009})}\BibitemShut
  {NoStop}%
\bibitem [{\citenamefont {Aichhorn}\ \emph {et~al.}(2011)\citenamefont
  {Aichhorn}, \citenamefont {Pourovskii},\ and\ \citenamefont
  {Georges}}]{Aichhorn2011}%
  \BibitemOpen
  \bibfield  {author} {\bibinfo {author} {\bibfnamefont {M.}~\bibnamefont
  {Aichhorn}}, \bibinfo {author} {\bibfnamefont {L.}~\bibnamefont
  {Pourovskii}}, \ and\ \bibinfo {author} {\bibfnamefont {A.}~\bibnamefont
  {Georges}},\ }\href@noop {} {\bibfield  {journal} {\bibinfo  {journal} {Phys.
  Rev. B}\ }\textbf {\bibinfo {volume} {84}},\ \bibinfo {pages} {054529}
  (\bibinfo {year} {2011})}\BibitemShut {NoStop}%
\bibitem [{\citenamefont {Blaha}\ \emph {et~al.}(2018)\citenamefont {Blaha},
  \citenamefont {Schwarz}, \citenamefont {Madsen}, \citenamefont {Kvasnicka},
  \citenamefont {Luitz}, \citenamefont {Laskowski}, \citenamefont {Tran},\ and\
  \citenamefont {Marks}}]{Wien2k}%
  \BibitemOpen
  \bibfield  {author} {\bibinfo {author} {\bibfnamefont {P.}~\bibnamefont
  {Blaha}}, \bibinfo {author} {\bibfnamefont {K.}~\bibnamefont {Schwarz}},
  \bibinfo {author} {\bibfnamefont {G.}~\bibnamefont {Madsen}}, \bibinfo
  {author} {\bibfnamefont {D.}~\bibnamefont {Kvasnicka}}, \bibinfo {author}
  {\bibfnamefont {J.}~\bibnamefont {Luitz}}, \bibinfo {author} {\bibfnamefont
  {R.}~\bibnamefont {Laskowski}}, \bibinfo {author} {\bibfnamefont
  {F.}~\bibnamefont {Tran}}, \ and\ \bibinfo {author} {\bibfnamefont {L.~D.}\
  \bibnamefont {Marks}},\ }\href@noop {} {\emph {\bibinfo {title} {WIEN2k, An
  augmented Plane Wave + Local Orbitals Program for Calculating Crystal
  Properties}}}\ (\bibinfo  {publisher} {Karlheinz Schwarz, Techn. Universität
  Wien, Austria,ISBN 3-9501031-1-2},\ \bibinfo {year} {2018})\BibitemShut
  {NoStop}%
\bibitem [{\citenamefont {Parcollet}\ \emph {et~al.}(2015)\citenamefont
  {Parcollet}, \citenamefont {Ferrero}, \citenamefont {Ayral}, \citenamefont
  {Hafermann}, \citenamefont {Krivenko}, \citenamefont {Messio},\ and\
  \citenamefont {Seth}}]{Parcollet2015}%
  \BibitemOpen
  \bibfield  {author} {\bibinfo {author} {\bibfnamefont {O.}~\bibnamefont
  {Parcollet}}, \bibinfo {author} {\bibfnamefont {M.}~\bibnamefont {Ferrero}},
  \bibinfo {author} {\bibfnamefont {T.}~\bibnamefont {Ayral}}, \bibinfo
  {author} {\bibfnamefont {H.}~\bibnamefont {Hafermann}}, \bibinfo {author}
  {\bibfnamefont {I.}~\bibnamefont {Krivenko}}, \bibinfo {author}
  {\bibfnamefont {L.}~\bibnamefont {Messio}}, \ and\ \bibinfo {author}
  {\bibfnamefont {P.}~\bibnamefont {Seth}},\ }\href {http://ipht.cea.fr/triqs/}
  {\bibfield  {journal} {\bibinfo  {journal} {Computer Physics Communications}\
  }\textbf {\bibinfo {volume} {196}},\ \bibinfo {pages} {398 } (\bibinfo {year}
  {2015})}\BibitemShut {NoStop}%
\bibitem [{\citenamefont {Aichhorn}\ \emph {et~al.}(2016)\citenamefont
  {Aichhorn}, \citenamefont {Pourovskii}, \citenamefont {Seth}, \citenamefont
  {Vildosola}, \citenamefont {Zingl}, \citenamefont {Peil}, \citenamefont
  {Deng}, \citenamefont {Mravlje}, \citenamefont {Kraberger}, \citenamefont
  {Martins} \emph {et~al.}}]{Aichhorn2016}%
  \BibitemOpen
  \bibfield  {author} {\bibinfo {author} {\bibfnamefont {M.}~\bibnamefont
  {Aichhorn}}, \bibinfo {author} {\bibfnamefont {L.}~\bibnamefont
  {Pourovskii}}, \bibinfo {author} {\bibfnamefont {P.}~\bibnamefont {Seth}},
  \bibinfo {author} {\bibfnamefont {V.}~\bibnamefont {Vildosola}}, \bibinfo
  {author} {\bibfnamefont {M.}~\bibnamefont {Zingl}}, \bibinfo {author}
  {\bibfnamefont {O.~E.}\ \bibnamefont {Peil}}, \bibinfo {author}
  {\bibfnamefont {X.}~\bibnamefont {Deng}}, \bibinfo {author} {\bibfnamefont
  {J.}~\bibnamefont {Mravlje}}, \bibinfo {author} {\bibfnamefont {G.~J.}\
  \bibnamefont {Kraberger}}, \bibinfo {author} {\bibfnamefont {C.}~\bibnamefont
  {Martins}},  \emph {et~al.},\ }\href@noop {} {\bibfield  {journal} {\bibinfo
  {journal} {Computer Physics Communications}\ }\textbf {\bibinfo {volume}
  {204}},\ \bibinfo {pages} {200} (\bibinfo {year} {2016})}\BibitemShut
  {NoStop}%
\bibitem [{\citenamefont {Amadon}\ \emph {et~al.}(2008)\citenamefont {Amadon},
  \citenamefont {Lechermann}, \citenamefont {Georges}, \citenamefont {Jollet},
  \citenamefont {Wehling},\ and\ \citenamefont {Lichtenstein}}]{Amadon2008}%
  \BibitemOpen
  \bibfield  {author} {\bibinfo {author} {\bibfnamefont {B.}~\bibnamefont
  {Amadon}}, \bibinfo {author} {\bibfnamefont {F.}~\bibnamefont {Lechermann}},
  \bibinfo {author} {\bibfnamefont {A.}~\bibnamefont {Georges}}, \bibinfo
  {author} {\bibfnamefont {F.}~\bibnamefont {Jollet}}, \bibinfo {author}
  {\bibfnamefont {T.~O.}\ \bibnamefont {Wehling}}, \ and\ \bibinfo {author}
  {\bibfnamefont {A.~I.}\ \bibnamefont {Lichtenstein}},\ }\href@noop {}
  {\bibfield  {journal} {\bibinfo  {journal} {Phys. Rev. B}\ }\textbf {\bibinfo
  {volume} {77}},\ \bibinfo {pages} {205112} (\bibinfo {year}
  {2008})}\BibitemShut {NoStop}%
\bibitem [{\citenamefont {Hosoi}\ \emph {et~al.}()\citenamefont {Hosoi},
  \citenamefont {Mizoguchi}, \citenamefont {Hinokihara}, \citenamefont
  {Hiroyasu~Matsuura},\ and\ \citenamefont {Ogata}}]{Hosoi}%
  \BibitemOpen
  \bibfield  {author} {\bibinfo {author} {\bibfnamefont {M.}~\bibnamefont
  {Hosoi}}, \bibinfo {author} {\bibfnamefont {T.}~\bibnamefont {Mizoguchi}},
  \bibinfo {author} {\bibfnamefont {T.}~\bibnamefont {Hinokihara}}, \bibinfo
  {author} {\bibfnamefont {H.}~\bibnamefont {Hiroyasu~Matsuura}}, \ and\
  \bibinfo {author} {\bibfnamefont {M.}~\bibnamefont {Ogata}},\ }\href@noop {}
  {\bibinfo  {journal} {arXiv:1804.04874 [cond-mat.str-el]}\ }\BibitemShut
  {NoStop}%
\bibitem [{\citenamefont {Arima}\ \emph {et~al.}(1969)\citenamefont {Arima},
  \citenamefont {Harvey},\ and\ \citenamefont {Shimizu}}]{ARIMA1969}%
  \BibitemOpen
\bibfield  {journal} {  }\bibfield  {author} {\bibinfo {author} {\bibfnamefont
  {A.}~\bibnamefont {Arima}}, \bibinfo {author} {\bibfnamefont
  {M.}~\bibnamefont {Harvey}}, \ and\ \bibinfo {author} {\bibfnamefont
  {K.}~\bibnamefont {Shimizu}},\ }\href {\doibase
  https://doi.org/10.1016/0370-2693(69)90443-2} {\bibfield  {journal} {\bibinfo
   {journal} {Physics Letters B}\ }\textbf {\bibinfo {volume} {30}},\ \bibinfo
  {pages} {517 } (\bibinfo {year} {1969})}\BibitemShut {NoStop}%
\bibitem [{\citenamefont {Santini}\ \emph {et~al.}(2009)\citenamefont
  {Santini}, \citenamefont {Carretta}, \citenamefont {Amoretti}, \citenamefont
  {Caciuffo}, \citenamefont {Magnani},\ and\ \citenamefont
  {Lander}}]{RevModPhys.81.807}%
  \BibitemOpen
  \bibfield  {author} {\bibinfo {author} {\bibfnamefont {P.}~\bibnamefont
  {Santini}}, \bibinfo {author} {\bibfnamefont {S.}~\bibnamefont {Carretta}},
  \bibinfo {author} {\bibfnamefont {G.}~\bibnamefont {Amoretti}}, \bibinfo
  {author} {\bibfnamefont {R.}~\bibnamefont {Caciuffo}}, \bibinfo {author}
  {\bibfnamefont {N.}~\bibnamefont {Magnani}}, \ and\ \bibinfo {author}
  {\bibfnamefont {G.~H.}\ \bibnamefont {Lander}},\ }\href {\doibase
  10.1103/RevModPhys.81.807} {\bibfield  {journal} {\bibinfo  {journal} {Rev.
  Mod. Phys.}\ }\textbf {\bibinfo {volume} {81}},\ \bibinfo {pages} {807}
  (\bibinfo {year} {2009})}\BibitemShut {NoStop}%
\bibitem [{\citenamefont {Shiina}\ \emph {et~al.}(1998)\citenamefont {Shiina},
  \citenamefont {Sakai}, \citenamefont {Shiba},\ and\ \citenamefont
  {Thalmeier}}]{doi:10.1143/JPSJ.67.941}%
  \BibitemOpen
  \bibfield  {author} {\bibinfo {author} {\bibfnamefont {R.}~\bibnamefont
  {Shiina}}, \bibinfo {author} {\bibfnamefont {O.}~\bibnamefont {Sakai}},
  \bibinfo {author} {\bibfnamefont {H.}~\bibnamefont {Shiba}}, \ and\ \bibinfo
  {author} {\bibfnamefont {P.}~\bibnamefont {Thalmeier}},\ }\href {\doibase
  10.1143/JPSJ.67.941} {\bibfield  {journal} {\bibinfo  {journal} {Journal of
  the Physical Society of Japan}\ }\textbf {\bibinfo {volume} {67}},\ \bibinfo
  {pages} {941} (\bibinfo {year} {1998})},\ \Eprint
  {http://arxiv.org/abs/https://doi.org/10.1143/JPSJ.67.941}
  {https://doi.org/10.1143/JPSJ.67.941} \BibitemShut {NoStop}%
\bibitem [{\citenamefont {Maharaj}\ \emph {et~al.}(2020)\citenamefont
  {Maharaj}, \citenamefont {Sala}, \citenamefont {Stone}, \citenamefont
  {Kermarrec}, \citenamefont {Ritter}, \citenamefont {Fauth}, \citenamefont
  {Marjerrison}, \citenamefont {Greedan}, \citenamefont {Paramekanti},\ and\
  \citenamefont {Gaulin}}]{PhysRevLett.124.087206}%
  \BibitemOpen
  \bibfield  {author} {\bibinfo {author} {\bibfnamefont {D.~D.}\ \bibnamefont
  {Maharaj}}, \bibinfo {author} {\bibfnamefont {G.}~\bibnamefont {Sala}},
  \bibinfo {author} {\bibfnamefont {M.~B.}\ \bibnamefont {Stone}}, \bibinfo
  {author} {\bibfnamefont {E.}~\bibnamefont {Kermarrec}}, \bibinfo {author}
  {\bibfnamefont {C.}~\bibnamefont {Ritter}}, \bibinfo {author} {\bibfnamefont
  {F.}~\bibnamefont {Fauth}}, \bibinfo {author} {\bibfnamefont {C.~A.}\
  \bibnamefont {Marjerrison}}, \bibinfo {author} {\bibfnamefont {J.~E.}\
  \bibnamefont {Greedan}}, \bibinfo {author} {\bibfnamefont {A.}~\bibnamefont
  {Paramekanti}}, \ and\ \bibinfo {author} {\bibfnamefont {B.~D.}\ \bibnamefont
  {Gaulin}},\ }\href {\doibase 10.1103/PhysRevLett.124.087206} {\bibfield
  {journal} {\bibinfo  {journal} {Phys. Rev. Lett.}\ }\textbf {\bibinfo
  {volume} {124}},\ \bibinfo {pages} {087206} (\bibinfo {year}
  {2020})}\BibitemShut {NoStop}%
\bibitem [{\citenamefont {Paramekanti}\ \emph {et~al.}(2020)\citenamefont
  {Paramekanti}, \citenamefont {Maharaj},\ and\ \citenamefont
  {Gaulin}}]{PhysRevB.101.054439}%
  \BibitemOpen
  \bibfield  {author} {\bibinfo {author} {\bibfnamefont {A.}~\bibnamefont
  {Paramekanti}}, \bibinfo {author} {\bibfnamefont {D.~D.}\ \bibnamefont
  {Maharaj}}, \ and\ \bibinfo {author} {\bibfnamefont {B.~D.}\ \bibnamefont
  {Gaulin}},\ }\href {\doibase 10.1103/PhysRevB.101.054439} {\bibfield
  {journal} {\bibinfo  {journal} {Phys. Rev. B}\ }\textbf {\bibinfo {volume}
  {101}},\ \bibinfo {pages} {054439} (\bibinfo {year} {2020})}\BibitemShut
  {NoStop}%
\bibitem [{\citenamefont {Winter}\ \emph {et~al.}(2016)\citenamefont {Winter},
  \citenamefont {Li}, \citenamefont {Jeschke},\ and\ \citenamefont
  {Valent\'{\i}}}]{PhysRevB.93.214431}%
  \BibitemOpen
  \bibfield  {author} {\bibinfo {author} {\bibfnamefont {S.~M.}\ \bibnamefont
  {Winter}}, \bibinfo {author} {\bibfnamefont {Y.}~\bibnamefont {Li}}, \bibinfo
  {author} {\bibfnamefont {H.~O.}\ \bibnamefont {Jeschke}}, \ and\ \bibinfo
  {author} {\bibfnamefont {R.}~\bibnamefont {Valent\'{\i}}},\ }\href {\doibase
  10.1103/PhysRevB.93.214431} {\bibfield  {journal} {\bibinfo  {journal} {Phys.
  Rev. B}\ }\textbf {\bibinfo {volume} {93}},\ \bibinfo {pages} {214431}
  (\bibinfo {year} {2016})}\BibitemShut {NoStop}%
\bibitem [{\citenamefont {Kim}\ \emph {et~al.}(2019)\citenamefont {Kim},
  \citenamefont {Efremov},\ and\ \citenamefont {van~den
  Brink}}]{PhysRevMaterials.3.014414}%
  \BibitemOpen
  \bibfield  {author} {\bibinfo {author} {\bibfnamefont {B.~H.}\ \bibnamefont
  {Kim}}, \bibinfo {author} {\bibfnamefont {D.~V.}\ \bibnamefont {Efremov}}, \
  and\ \bibinfo {author} {\bibfnamefont {J.}~\bibnamefont {van~den Brink}},\
  }\href {\doibase 10.1103/PhysRevMaterials.3.014414} {\bibfield  {journal}
  {\bibinfo  {journal} {Phys. Rev. Materials}\ }\textbf {\bibinfo {volume}
  {3}},\ \bibinfo {pages} {014414} (\bibinfo {year} {2019})}\BibitemShut
  {NoStop}%
\bibitem [{\citenamefont {T.~Inui}\ and\ \citenamefont
  {Onodera}(1990)}]{Inui1990}%
  \BibitemOpen
  \bibfield  {author} {\bibinfo {author} {\bibfnamefont {Y.~T.}\ \bibnamefont
  {T.~Inui}}\ and\ \bibinfo {author} {\bibfnamefont {Y.}~\bibnamefont
  {Onodera}},\ }\href {\doibase 10.1007/978-3-642-80021-4} {\emph {\bibinfo
  {title} {Group theory and its applications in physics}}}\ (\bibinfo
  {publisher} {Springer-Verlag Berlin Heidelberg},\ \bibinfo {year}
  {1990})\BibitemShut {NoStop}%
\bibitem [{\citenamefont {Horvat}\ \emph {et~al.}(2017)\citenamefont {Horvat},
  \citenamefont {Pourovskii}, \citenamefont {Aichhorn},\ and\ \citenamefont
  {Mravlje}}]{PhysRevB.95.205115}%
  \BibitemOpen
  \bibfield  {author} {\bibinfo {author} {\bibfnamefont {A.}~\bibnamefont
  {Horvat}}, \bibinfo {author} {\bibfnamefont {L.}~\bibnamefont {Pourovskii}},
  \bibinfo {author} {\bibfnamefont {M.}~\bibnamefont {Aichhorn}}, \ and\
  \bibinfo {author} {\bibfnamefont {J.}~\bibnamefont {Mravlje}},\ }\href
  {\doibase 10.1103/PhysRevB.95.205115} {\bibfield  {journal} {\bibinfo
  {journal} {Phys. Rev. B}\ }\textbf {\bibinfo {volume} {95}},\ \bibinfo
  {pages} {205115} (\bibinfo {year} {2017})}\BibitemShut {NoStop}%
\bibitem [{\citenamefont {Franchini}\ \emph {et~al.}(2011)\citenamefont
  {Franchini}, \citenamefont {Archer}, \citenamefont {He}, \citenamefont
  {Chen}, \citenamefont {Filippetti},\ and\ \citenamefont
  {Sanvito}}]{PhysRevB.83.220402}%
  \BibitemOpen
  \bibfield  {author} {\bibinfo {author} {\bibfnamefont {C.}~\bibnamefont
  {Franchini}}, \bibinfo {author} {\bibfnamefont {T.}~\bibnamefont {Archer}},
  \bibinfo {author} {\bibfnamefont {J.}~\bibnamefont {He}}, \bibinfo {author}
  {\bibfnamefont {X.-Q.}\ \bibnamefont {Chen}}, \bibinfo {author}
  {\bibfnamefont {A.}~\bibnamefont {Filippetti}}, \ and\ \bibinfo {author}
  {\bibfnamefont {S.}~\bibnamefont {Sanvito}},\ }\href {\doibase
  10.1103/PhysRevB.83.220402} {\bibfield  {journal} {\bibinfo  {journal} {Phys.
  Rev. B}\ }\textbf {\bibinfo {volume} {83}},\ \bibinfo {pages} {220402}
  (\bibinfo {year} {2011})}\BibitemShut {NoStop}%
\bibitem [{\citenamefont {Archer}\ \emph {et~al.}(2011)\citenamefont {Archer},
  \citenamefont {Pemmaraju}, \citenamefont {Sanvito}, \citenamefont
  {Franchini}, \citenamefont {He}, \citenamefont {Filippetti}, \citenamefont
  {Delugas}, \citenamefont {Puggioni}, \citenamefont {Fiorentini},
  \citenamefont {Tiwari},\ and\ \citenamefont {Majumdar}}]{PhysRevB.84.115114}%
  \BibitemOpen
  \bibfield  {author} {\bibinfo {author} {\bibfnamefont {T.}~\bibnamefont
  {Archer}}, \bibinfo {author} {\bibfnamefont {C.~D.}\ \bibnamefont
  {Pemmaraju}}, \bibinfo {author} {\bibfnamefont {S.}~\bibnamefont {Sanvito}},
  \bibinfo {author} {\bibfnamefont {C.}~\bibnamefont {Franchini}}, \bibinfo
  {author} {\bibfnamefont {J.}~\bibnamefont {He}}, \bibinfo {author}
  {\bibfnamefont {A.}~\bibnamefont {Filippetti}}, \bibinfo {author}
  {\bibfnamefont {P.}~\bibnamefont {Delugas}}, \bibinfo {author} {\bibfnamefont
  {D.}~\bibnamefont {Puggioni}}, \bibinfo {author} {\bibfnamefont
  {V.}~\bibnamefont {Fiorentini}}, \bibinfo {author} {\bibfnamefont
  {R.}~\bibnamefont {Tiwari}}, \ and\ \bibinfo {author} {\bibfnamefont
  {P.}~\bibnamefont {Majumdar}},\ }\href {\doibase 10.1103/PhysRevB.84.115114}
  {\bibfield  {journal} {\bibinfo  {journal} {Phys. Rev. B}\ }\textbf {\bibinfo
  {volume} {84}},\ \bibinfo {pages} {115114} (\bibinfo {year}
  {2011})}\BibitemShut {NoStop}%
\bibitem [{leo()}]{leonid}%
  \BibitemOpen
  \href@noop {} {}\bibinfo {note} {L. V. Pourovskii, private
  communication}\BibitemShut {NoStop}%
\bibitem [{\citenamefont {Pourovskii}\ and\ \citenamefont
  {Khmelevskyi}(2019)}]{PhysRevB.99.094439}%
  \BibitemOpen
  \bibfield  {author} {\bibinfo {author} {\bibfnamefont {L.~V.}\ \bibnamefont
  {Pourovskii}}\ and\ \bibinfo {author} {\bibfnamefont {S.}~\bibnamefont
  {Khmelevskyi}},\ }\href {\doibase 10.1103/PhysRevB.99.094439} {\bibfield
  {journal} {\bibinfo  {journal} {Phys. Rev. B}\ }\textbf {\bibinfo {volume}
  {99}},\ \bibinfo {pages} {094439} (\bibinfo {year} {2019})}\BibitemShut
  {NoStop}%
\end{thebibliography}%

\clearpage

\section{Supplementary Material}

\section{Methods}

In this section the computational approaches used to investigate the multipolar interactions are described.

\subsection{Computational details}

Density Functional Theory (DFT) calculations were done within the 
projector augmented wave method~\cite{PhysRevB.50.17953} as implemented in the Vienna ab initio Simulation Package (VASP)~\cite{PhysRevB.47.558, PhysRevB.54.11169}. 
The Perdew-Burke-Ernzerhof approximation~\cite{PhysRevLett.77.3865} for the exchange-correlation functional was  used, and, in order to take into account the relativistic effects, we  considered the fully relativistic scheme that includes Spin Orbit Coupling (SOC), with a SOC energy of 482 meV (computed by mapping the DFT energy difference calculated with and without SOC onto a relativistic atomic Hamiltonian for $d$ orbitals~\cite{PhysRevMaterials.3.083802}). 
Furthermore, we added an on-site Hubbard U correction as implemented in the Dudarev's approach for noncollinear spin configurations~\cite{PhysRevMaterials.3.083802}:
\begin{equation}
       E_{DFT+U} = E_{DFT} + \frac{U}{2} \sum_{\sigma} \Big[ \Big( \sum_{m_{1}} n^{\sigma}_{m_{1}, m_{1}} \Big) - \Big( \sum_{m_{1},m_{2}} \hat{n}^{\sigma}_{m_{1},m_{2}} \hat{n}^{\sigma}_{m_{2},m_{1}} \Big) \Big] \ , 
\end{equation}
where $\hat{n}^{\sigma}_{m_{1},m_{2}}$ is density matrix with $m_{1}$ and $m_{2}$ running over the orbital angular momentum and $\sigma$ over the spin angular momentum and U is the effective Hubbard U interaction.
We used a U of  3.4 eV, in line with previous experimental estimations ~\cite{PhysRevLett.99.016404}. As a reference we show in Fig.~\ref{fig:spectra} the optical spectra obtained with this value of $U$ at DFT+U+SOC and ED level.

\begin{figure*}[h]
   \begin{center}
   \includegraphics[width=0.6\textwidth,clip=true]{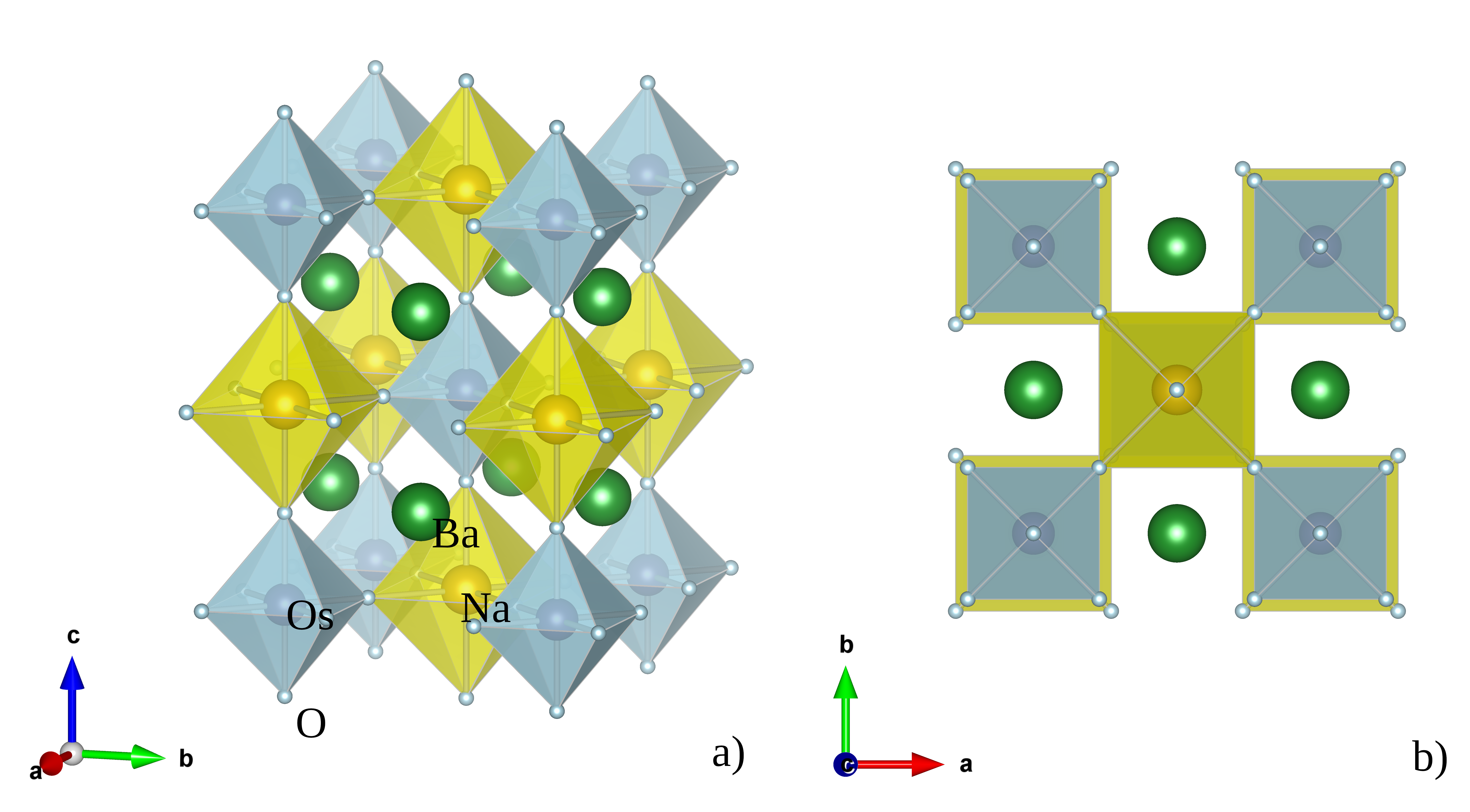}
    \end{center}
\caption{Magnetic unit cell of BNOO.}
\label{fig:unit_cell}
\end{figure*}

The supercell used is shown in Figure ~\ref{fig:unit_cell}, where the lattice parameters are $a/\sqrt{2} \times a/\sqrt{2} \times a$, with $a$ = 8.287 \AA $ $ being the experimental room-temperature value ~\cite{STITZER2002311}. It contains two formula units, or, equivalently, two osmium atoms on parallel planes along the z direction.
The reciprocal space was sampled with a $k$-point mesh of 6$\times$6$\times$6 and a high accuracy was achieved by setting the energy cutoff to the plane waves of 600 eV, together with a electronic convergence criterion of $\sim$ 10$^{-8}$ eV.

The local structure was studied by means of selective dynamics routine, fixing the positions of the Na, Os and Ba sites and allowing the oxygens to be relaxed to their ground state through the  direct inversion in the iterative subspace (RMM-DIIS) algorithm.
The magnetic configuration in the canted phase was set with noncollinear magnetic moments constrained as in Figure 1 (b) (main text), with $\phi = 67^{o}$. 
Both DFT, DFT + U  and DFT + U + SOC were used in order to probe the role of the different degrees of freedom. 
We verified that the most stable local structure is the one defined as Model A in Ref. ~\cite{PhysRevB.97.224103}.

\begin{figure*}[h]
   \begin{center}
   \includegraphics[width=0.6\textwidth,clip=true]{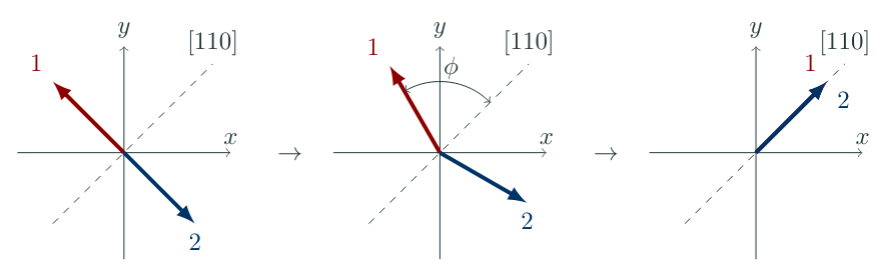}
    \end{center}
\caption{Canting angles for BNOO}
\label{fig:canting_1}
\end{figure*}

The canted procedure used for studying the exchange interactions -- keeping the structure fixed and changing the direction of the local magnetic moments -- allows to make use of the magnetic force theorem.
To do so, we tilted  the angles between magnetic moments in parallel planes in steps of 2.5 degree starting from a AFM [$\bar{1}$10] and ending in the FM [110] configuration (see Figure ~\ref{fig:canting_1})

This can be achieved in VASP by means of the constrained magnetic moment approach which constrains the direction of the local spin magnetic moment by adding an extra penalty energy, which has the following expression~\cite{Liu2015}
\begin{equation}
       E(\lambda) = \sum_{I} \lambda \Big[ \mathbf{\vec{M}}_{I} - \mathbf{\hat{M}}_{I}^{0} \big(\mathbf{\hat{M}}_{I}^{0} \cdot \mathbf{M}_{I} \big) \Big]^{2}  \ .
\end{equation}
Here the sum runs over the ionic sites $I$, $\mathbf{\hat{M}}_{I}^{0}$ is the unit vector that gives the constrained direction of the magnetic moment at site $ I$, and $\vec{M}_I$ is the integrated magnetic moment inside the Muffin-tin sphere $\Omega_I$. 
The constant $\lambda$ gives the strength of the penalty energy and was chosen manually in order to get a value of $E(\lambda) \ll E(DFT)$, a necessary requirement for having meaningful DFT total energies. 
Our values of $E(\lambda)$ were always lower than 10$^{-6}$ eV.

The result  is a energy variation as a function of the canting angle, that, as we will show in the next section, can be mapped onto a Pseudo Spin Hamiltonian (PSH) with multipolar interactions.

\subsection{Calculations of exchange interactions within DFT+HI }

Our charge self-consistent DFT+DMFT calculations using the Hubbard-I (HI) approximation for Os 5$d$, abbriviated as DFT+HI, were carried out for the paramagnetic phase and  the undistored cubic structure of BNOO with $a=8.287$~\AA\ and the Os-O distance  of 1.869~\AA. We employed the Wien-2k electronic structure package in conjunction with "TRIQS" library implementations for the DMFT cycle and  Hubbard-I method. The spin-orbit coupling was included in Wien2k within  the standard second-variation treatment. The Brillouin zone (BZ) integration was carried out using 400~{\bf k}-points in the full BZ.  

The Wannier orbitals representing $t_{2g}$  states of Os 5$d$ were constructed by the projective technique of Refs.~\cite{Amadon2008,Aichhorn2009} using the bands enclosed by the energy window $[-1.09:2.04]$~eV; this window thus encloses all $t_{2g}$-like bands. The Coulomb interaction was specified by the Slater parameter $F_0=$3.2~eV and the Hund's rule coupling $J_H$=0.5~eV. We employed the rotationally invariant Kanamori Coulomb repulsion between Os $t_{2g}$, which was correspondingly  defined by the intra-orbital repulsion $U_K=F_0+4J_H/5$ and the Kanamori Hund's rule $J_K=3J_H/5$. The double counting was evaluated using the fully-localized-limit expression and the atomic occupancy $N=1$ of Os 5$d$ shell.  The DFT+HI calculations were converged to 0.01 mRy in the total energy.

The DFT+HI calculations for the paramagnetic BNOO phase predict the  ground state quadruplet $J_{eff}=3/2$ for the Os 5$d^1$ shell, as expected; the excited doublet $J_{eff}=1/2$ ws predicted to be  0.45~eV higher in energy, corresponding to the spin-orbit coupling strength  $\lambda$ for the $t_{2g}$ subshell equal to 0.3~eV. We subsequently employed the FT-HI method of Ref.~\cite{PhysRevB.94.115117} to evaluate all nearest-neighbour exchange interactions between the $J_{eff}=3/2$ ground state multiplets. Namely, matrix elements of inter-site coupling $V^{\bf RR'}$ between sites $\vR$ and $\vR'$ read:
	\begin{equation}\label{V}
 \langle M_1 M_3| V^{\bf RR'}| M_2 M_4\rangle=\mathrm{Tr} \left[ G_{\bf RR'}\frac{\delta\Sigma^{at}_{\bf R'}}{\delta \rho^{M_3M_4}_{\vR'}} G_{\bf R'R}\frac{\delta\Sigma^{at}_{\bf R}}{\delta \rho^{M_1M_2}_{\vR}}\right],
	\end{equation}
where $M$ is the projection quantum number, $M=-3/2...3/2$, for the multiplet $J_{eff}=3/2$, $\rho^{M_iM_j}_{\vR}$ is the corresponding element of the $J_{eff}=3/2$ density matrix on site $R$, $\frac{\delta\Sigma^{at}_{\bf R}}{\delta \rho^{M_iM_j}_{\vR}}$ is the derivative of atomic (Hubbard-I) self-energy $\Sigma^{at}_{\bf R}$ over a fluctuation of the $\rho^{M_iM_j}_{\vR}$ element, $G_{\bf RR'}$ is the inter-site Green's function evaluated within the DFT+HI.  Once all matrix elements (\ref{V}) are calculated, they are transformed to the couplings $I_{KK'}^{QQ'}$ between on-site moments (eq.~1 of the main text) as follows:
$$
I^{QQ'}_{KK'}({\bf RR'})=\sum_{\substack{M_1M_2 \\ M_3M_4}} \langle M_1M_3| V^{\bf RR'}|  M_2M_4\rangle \left[O_{Q}^{K}(J)\right]_{M_2M_1} \left[O_{Q'}^{K'}(J)\right]_{M_4M_3},
$$
where $\left[O_{Q}^{K}(J)\right]_{M_1M_2}$ is the $M_1M_2$ matrix element of the corresponding multipolar tensor.

The resulting 15$\times$15 interaction matrix $I^{QQ'}_{KK'}$ for the [1/2,1/2,0] Os-Os bond (omitting the irrelevant monopole term) is shown in Table~\ref{table:V}. Interaction matrices for other NN bonds can be obtained from it using the corresponding symmetry operations;  the interaction matrix summed over all interlayer bonds and omitting the terms irrelevant for the planar magnetic order of BNOO is graphically depicted in Fig.~2c  of the main text. 

\renewcommand{\arraystretch}{1.5}
\begin{table*}
\setlength{\tabcolsep}{6pt}
	\caption{\label{table:V} The interaction matrix $I^{QQ'}_{KK'}$ calculated by DFT+HI for the nearest-neighbour Os-Os bond [1/2,1/2,0], in meV. The order of $K$ and $Q$ is indicated in the first two columns.}
	\centering
	\begin{tabular}{rr|rrr|rrrrr|rrrrrrr}
      &  -1  & -0.96 & -0.03 & -0.74 & 0 & 0 & 0 & 0 & 0 & 0.14 & -0.01 & -1.05 & 0 & 0.19 & 0 & -0.54  \\ 
 1  &  0  & -0.03 & 1.36 & -0.02 & 0 & 0 & 0 & 0 & 0 & 0 & -0.83 & -0.01 & 0.44 & 0 & 0.01 & 0  \\ 
     &  1  & -0.74 & -0.02 & -0.85 & 0 & 0 & 0 & 0 & 0 & 0.52 & 0 & 0.17 & 0 & -1.05 & 0 & -0.09  \\ 
     \hline
     &  -2  & 0 & 0 & 0 & -1.04 & 0 & -1.02 & -0.02 & -0.01 & 0 & 0 & 0 & 0 & 0 & 0 & 0  \\ 
     &  -1  & 0 & 0 & 0 & 0 & -1.65 & 0.01 & 0.10 & -0.01 & 0 & 0 & 0 & 0 & 0 & 0 & 0  \\ 
 2  &  0  & 0 & 0 & 0 & -1.02 & 0.01 & 1.43 & 0.01 & 0.07 & 0 & 0 & 0 & 0 & 0 & 0 & 0  \\ 
     &  1  & 0 & 0 & 0 & -0.02 & 0.10 & 0.01 & -1.76 & 0.02 & 0 & 0 & 0 & 0 & 0 & 0 & 0  \\ 
     &  2  & 0 & 0 & 0 & -0.01 & -0.01 & 0.07 & 0.02 & -1.42 & 0 & 0 & 0 & 0 & 0 & 0 & 0  \\ 
        \hline
     &  -3  & 0.14 & 0 & 0.52 & 0 & 0 & 0 & 0 & 0 & 1.69 & -0.01 & -0.20 & 0.01 & 0.49 & -0.01 & 0  \\ 
     &  -2  & -0.01 & -0.83 & 0 & 0 & 0 & 0 & 0 & 0 & -0.01 & -1.14 & 0 & 0.32 & 0.02 & 0 & -0.02  \\ 
     &  -1  & -1.05 & -0.01 & 0.17 & 0 & 0 & 0 & 0 & 0 & -0.20 & 0 & 0.19 & -0.01 & 0.58 & 0.01 & -0.49  \\ 
 3  &  0  & 0 & 0.44 & 0 & 0 & 0 & 0 & 0 & 0 & 0.01 & 0.32 & -0.01 & 1.34 & 0 & -0.09 & 0  \\ 
     &  1  & 0.19 & 0 & -1.05 & 0 & 0 & 0 & 0 & 0 & 0.49 & 0.02 & 0.58 & 0 & 0.32 & 0 & 0.08  \\ 
     &  2  & 0 & 0.01 & 0 & 0 & 0 & 0 & 0 & 0 & -0.01 & 0 & 0.01 & -0.09 & 0 & -1.49 & 0.01  \\ 
     &  3  & -0.54 & 0 & -0.09 & 0 & 0 & 0 & 0 & 0 & 0 & -0.02 & -0.49 & 0 & 0.08 & 0.01 & 1.68  \\ 
	\end{tabular}
	\label{tab:intermathi}
\end{table*}

\subsection{Exact Diagonalization}

To calculate the multipolar interactions between neighboring Os$^{7+}$ ions,
we considered two-site $t_{2g}$-orbital systems with following Hamiltonian
\begin{align}
\label{Eq_ED}
H &=  \sum_{i\mu\nu\sigma}  \epsilon_{\mu\nu}^i c^\dagger_{i\mu\sigma}c_{i\nu\sigma} +
   \lambda \sum_{i\mu\nu\sigma\sigma^{\prime}}
   (\mathbf{l}\cdot\mathbf{s})_{\mu\sigma,\nu\sigma^{\prime}}
   c_{i\mu\sigma}^{\dagger}c_{i\nu\sigma^{\prime}} \nonumber
  + \frac{1}{2}\sum_{i\sigma\sigma^{\prime}\mu\nu}
  U_{\mu\nu} c_{i\mu\sigma}^{\dagger}c_{i\nu\sigma^{\prime}}^{\dagger}
  c_{i\nu\sigma^{\prime}}c_{i\mu\sigma}
  + \frac{1}{2}\sum_{\substack{i\sigma \sigma^{\prime} \\ \mu\ne\nu}}
  J_{\mu\nu}c_{i\mu\sigma}^{\dagger}c_{i\nu\sigma^{\prime}}^{\dagger}
  c_{i\mu\sigma^{\prime}}c_{i\nu\sigma}  \nonumber \\
  &+ \frac{1}{2}\sum_{\substack{\sigma \\ \mu\ne\nu}}
  J_{\mu\nu}^{\prime} c_{i\mu\sigma}^{\dagger}c_{i\mu,-\sigma}^{\dagger}
  c_{i\nu,-\sigma}c_{i\nu\sigma}
 + \sum_{\mu\nu\sigma} \left(
  t_{\mu\nu}^{12}c^\dagger_{1\mu\sigma}c_{2\nu\sigma} + h.c. \right),
\end{align}
where $c^\dagger_{i\mu\sigma}$ is the creation operator of $t_{2g}$ electrons
 with $\mu$ ($\mu = xy,yz,zx$) orbital and $\sigma$ ($\sigma = +,-$) 
spin states at the $i$-the site ($i=1,2$).
First and second terms describe local level splitting of $t_{2g}$ orbitals
due to the crystal field and spin-orbit coupling, respectively.
The spin-orbit coupling strength $\lambda$ is set to be $0.3$ eV.
Third, fourth, and fifth terms refer to the Kanamori-type Coulomb interactions
with $U_{\mu\mu}=U$, $U_{\mu\ne\nu}=U-2J_H$, and $J_{\mu\nu}=J_{\mu\nu}^{\prime}=J_H$.
We set $U=3.4$ and $J_H=0.3$ eV.
Last term is the hopping Hamiltonian between $t_{2g}$ orbitals at 1 and 2 sites.
The hopping integral $t_{\mu\nu}^{12}$ and local crystal splitting 
$\epsilon_{\mu\nu}^i$ are estimated with the Wannier projection 
of the DFT calculation.
We took into account all possible states of two electrons in two-site cluster.
Employing the exact diagonalization method, we solved Eq.~\ref{Eq_ED}
and got eigenvalues and eigenstates.

Because $U$ is much larger than strength of hopping integrals,
lowet $16$ eigenstates are dominantly determined by 
unperturbative eigenstates $|M_1M_2\rangle$'s ($-3/2\le M_1, M_2 \le 3/2$).
Let $E_n$ and $\left| \Psi_n \right>$ be $n$-th eigenvalue and eigenstate, respectively. 
The spin-orbital dynamics of $J_{eff}=3/2$ multiplets can be
depicted with following effective Hamiltonian
\begin{equation}
H_{\rm eff} 
 = \sum_{M_1M_2M_1'M_2'}  
      h^{M_1M_2}_{M_1'M_2'}| M_1 M_2 \rangle \langle M_1'M_2' |.
\label{eq_Heff}
\end{equation}
The overlap matrix $\mathbf{S}$ is defined as
$S_{nm}=\sum_{M_1M_2}
\left< \Psi_n \right| M_1M_2 \left>\right< M_1M_2 \left|\Psi_m \right>$.
Coefficient $h^{M_1M_2}_{M_1'M_2'}$ is approximately estimated as
\begin{equation}
h^{M_1M_2}_{M_1'M_2'} = \sum_{n=1}^{16}\sum_{m=1}^{16}\sum_{m'=1}^{16}
E_n [S^{-1/2}]_{mn} [S^{-1/2}]_{nm'}
 \langle M_1M_2 |\Psi_m \rangle \langle \Psi_{m'} | M_1' M_2' \rangle,
\end{equation}
where $\mathbf{S}^{-1/2}$ is the inverse of square-root matrix of $\mathbf{S}$
\cite{PhysRevB.93.214431,PhysRevMaterials.3.014414}.

Let $T^K_Q(J)$ be the irreducible tensor operator in a $J$ manifold
($K=0,\cdots,2J$ and $Q=-K,\cdots,K$).
According to the Wigner-Eckart theorem, $T^K_Q(J)$ is given as
\begin{equation}
T^K_Q(J) = \sum_{MM'}
\langle J \vert | T^{K}(J) |\vert J \rangle 
\frac{ C_{J'M'KQ}^{JM}}{\sqrt{2J+1}} \left| J, M \right> \left< J, M'\right|,
\end{equation}
where,$\langle J \vert | T^{K}(J)|\vert J \rangle$
is the reduced matrix element of rank $K$ tensors and $C_{J'M'KQ}^{JM}$
is the Clebsch-Gordan (CG) coefficients  defined as
$C_{J'M'KQ}^{JM}= \langle J M \vert J'M' K Q \rangle$~\cite{Inui1990}.
Note that CG coefficients are satisfied with following orthogonality relation
\begin{equation}
\sum_{KQ} \frac{2K+1}{2J+1}
C^{JM_1}_{J M_2 K Q} C^{JM_1'}_{J M_2' K Q}
= \delta_{M_1M_2}\delta_{M'_1M'_2}.
\end{equation} 
We can easily show the following relation
\begin{equation}
\left| J, M \right> \left< J, M'\right| = \sum_{KQ} \frac{2K+1}{\sqrt{2J+1}}
\frac{C^{J M}_{J M' K Q}}{\langle J \vert | T^{K}(J) |\vert J \rangle} T^K_Q(J)
= \sum_{KQ} A^{JMM'}_{KQ}T^K_Q(J),
\end{equation}
where $A^{JMM'}_{KQ} = \frac{2K+1}{\sqrt{2J+1}}
\frac{C^{J M}_{J M' K Q}}{\langle J \vert | T^{K} |\vert J \rangle}$.
The multipolar tensor operator $O^K_Q(J)$ of real-valued (tesseral) harmonics
are built as following
\begin{subequations}
\begin{align}
 O^K_0(J) &= T^K_0(J)  \\
 O^K_Q(J) &= \frac{1}{\sqrt{2}} 
\Big( T^K_{-Q}(J) + (-1)^Q T^K_Q(J) \Big) ~(Q>0) \\
O^K_{-Q}(J) &= \frac{i}{\sqrt{2}} 
\Big( T^K_{-Q}(J) - (-1)^Q T^K_Q(J) \Big) ~(Q>0).
\end{align}
\end{subequations}
The projection operator $\left|J, M \right> \left<J, M'\right|$ is given as
\begin{equation}
\left| J,M \right> \left< J,M'\right| = \sum_{KQ} B^{JMM'}_{KQ}O^K_Q(J),
\end{equation}
where $B^{JMM'}_{K0}=A^{JMM'}_{K0}$, 
$B^{JMM'}_{KQ}= \frac{A^{JMM'}_{K,-Q}+(-1)^QA^{JMM'}_{KQ}}{\sqrt{2}}$,
and $B^{JMM'}_{K,-Q}= \frac{A^{JMM'}_{K,-Q}-(-1)^QA^{JMM'}_{KQ}}{i\sqrt{2}}$ ($Q>0$).
Accordingly, the effective Hamiltonian in Eq.~\ref{eq_Heff} 
is expressed in terms of $O^K_Q$ operators as following
\begin{equation}
H_{\rm eff} = I^{00}_{00} \mathbf{I}+ 
\sideset{}{'}\sum_{KQ} I^{K0}_{Q0} O^K_Q(J_1) + 
\sideset{}{'}\sum_{KQ} I^{0K}_{0Q} O^K_Q(J_2) + 
\sideset{}{'}\sum_{K_1Q_1}\sideset{}{'}\sum_{K_2Q_2} 
I^{K_1K_2}_{Q_1Q_2} O^{K_1}_{Q_1}(J_1)O^{K_2}_{Q_2}(J_2),
\end{equation}
where $I^{K_1K_2}_{Q_1Q_2} = \sum_{M_1M_2M'_1M'_2} h^{M_1M_2}_{M'_1M'_2}
B^{JM_1M'_1}_{K_1Q_1}B^{JM_2M'_2}_{K_2Q_2}$ and 
$\sideset{}{'}\sum_{KQ}$ refers to the summation over 
all $K$ and $Q$ values except for $K=Q=0$.

  \renewcommand{\arraystretch}{1.5}
\begin{table*}
\setlength{\tabcolsep}{6pt}
	\caption{\label{table:V} The interaction matrix $I^{K_{1}K_{2}}_{Q_{1}Q_{2}}$ calculated by ED for the nearest-neighbour Os-Os bond [1/2,1/2,0], in meV. The order of $K$ and $Q$ is indicated in the first two columns.}
	\centering
	\begin{tabular}{rr|rrr|rrrrr|rrrrrrr}
    &  -1 & -0.68  &  0 &   -0.61  &  0 &   0 &  0 & 0 & 0 &  0.06 &   0 &  -1.07  &  0 &  0.11 &  0 &  -0.28 \\ 
 1  & 0 & 0 & 0.90  & 0 & 0  &  0 &   0 &  0 &  0 &  0 &  -0.53  & 0 &  0.15 &  0 &  0  & 0  \\ 
     & 1 & -0.61  & 0 &  -0.68  & 0 &   0 & 0 &  0 & 0  & 0.28 &  0 &  0.11  & 0 &  -1.07  &  0 &  -0.06\\ 
     \hline
     & -2 & 0  &  0 & 0 &  -1.17  & 0 &   0.80  & 0 &  0  & 0 &  0 &  0 &  0 & 0 & 0 &  0 \\ 
     & -1 & 0 &  0 &  0 & 0 & -1.63 &  0 &  0.08 &  0 &  0 & 0 &  0 &  0 &  0 &  0 &  0 \\ 
 2  & 0 & 0 & 0 & 0 & 0.80  &  0 & 1.51  &  0 & 0 &  0 &  0 & 0 & 0 & 0 & 0 & 0\\ 
     & 1 & 0 &  0 & 0 &  0 &   0.08   &  0 & -1.63  &  0 &  0 & 0 & 0 & 0 & 0 & 0 &  0\\
     & 2 & 0 &   0 & 0 & 0 &  0 & 0 & 0 & -1.42  & 0 &  0 &  0 &  0 &   0 &  0 &  0\\ 
        \hline
     &  -3  & 0.06  &  0 & 0.28 & 0 & 0 &  0 &  0 & 0 & 1.09 &  0 & -0.07 &  0 &  0.28  & 0 & 0\\ 
     &  -2  & 0 & -0.53 &  0 & 0 & 0 & 0 &  0 & 0 &  0 & -1.08  & 0 &  0.31  & 0 &  0 &   0 \\ 
     &  -1  & -1.07  & 0 &   0.11  & 0 &  0 &  0 &   0 &  0 & -0.07 &  0 &  0.33  & 0 & 0.47  & 0 & -0.28\\ 
 3  &  0  & 0 &  0.15  &  0 &  0 &  0 &  0 &  0 &  0 &  0 &  0.31  & 0 &  1.30 &  0 & 0 & 0\\ 
     &  1  & 0.11 &  0 &  -1.07  & 0 & 0 & 0 & 0 &  0 & 0.28 &  0 & 0.47 &  0 &  0.33  & 0 &  0.07\\ 
     &  2  & 0 &  0 & 0 & 0 &  0 &  0 & 0 & 0 &  0 &  0 & 0 &  0 &  0 &  -1.26  &  0\\ 
     &  3  & -0.28  & 0 &  -0.06  &  0 &  0 &  0 &  0 &  0 & 0 &  0 &  -0.28 &  0 &   0.07  & 0 &   1.09\\ 
	\end{tabular}
	\label{tab:intermated}
\end{table*}

\section{Optimal Structural Data}
\label{sec:structure}

The energy minimum is found for an expansion of the Os-O octahedra along the x, y, and z directions with different magnitude along the different crystallographic axis. 
Between parallel planes along z, the distortions are staggered in the xy plane, meaning that the Os(1)-O bond length along x (y) is equal to the Os(2)-O bond length along y (x). 
Furthermore, the angles between bonds in the xy plane are slightly changed from the perfect 90 degrees, forming alternated angles of $\sim$ 90.5$^{o}$ and $\sim$ 89.5$^{o}$. 

The corresponding Point Group is D$_{2h-12}$ or Pmnn and allows for anti-symmetric exchange interactions. The structural CIF file for this compound is in Appendix.

\begin{table}[h]
\begin{center}
\begin{tabular}{| c | c c c |}
\hline
Atoms &  x & y & z\\
\hline
Na & 0	&  0	&  0 \\
 & 0.5	&  0.5	&  0.5 \\
\hline
Os & 0.5	&  0.5	&  0 \\
 & 0	&  0	&  0.5 \\
\hline
Ba & 0.5	&  0	&  0.75 \\
 & 0	&  0.5	&  0.75 \\
 & 0.5	&  0	&  0.25 \\
 & 0	&  0.5	&  0.25 \\
\hline
O & 0.27126	&  0.26896	&  0.0 \\
 & 0.72874	&  0.73104	&  0.0 \\
 & 0.72667	&  0.27112	&  0.0 \\
 & 0.27333	&  0.72888	&  0.0 \\
 & 0.22667	&  0.22888	&  0.5 \\
 & 0.77333	&  0.77112	&  0.5 \\
 & 0.77126	&  0.23104	&  0.5 \\
 & 0.22874	&  0.76896	&  0.5 \\ 
 & 0	&  0	&  0.72849 \\
 & 0	&  0	&  0.27151 \\
 & 0.5	&  0.5	&  0.77151 \\
 & 0.5	&  0.5	&  0.22849 \\
\hline
\end{tabular}
\caption{\label{tab:table_positions} Optimized structural positions in direct coordinates.}
\end{center}
\end{table}

\subsection{Jahn-Teller distortions}
The JT vibrational mode 
Q$_{1}$, Q$_{2}$ and Q$_{3}$, are defined following the standard Van Vleck notation~\cite{Vleck1939}: 

\begin{align}
    Q_{1} &= \frac{dx + dy + dz}{\sqrt{6}}\\
    Q_{2} &= \frac{dx - dy}{2}\\
    Q_{3} &= \frac{dx + dy - 2dz}{2\sqrt{3}} \\
    \theta &= \tan^{-1} \Big( \frac{Q_{2}}{Q_{3}} \Big)\\
    Q &= \sqrt{Q_{2}^{2} + Q_{3}^{2}} \ ,
\end{align}
where dx (dy, dz) is x$_{i}$ - x$_{0}$ (y$_{i}$ - y$_{0}$ , z$_{i}$ - z$_{0}$ )  i.e. the difference between the distorted Os-O bond length and the undistorted one. 

To construct the phase diagram we have introduced the relative JT parameter $\tilde{Q}_{1}$, defined as the ratio between Q$_{1}$ for the interpolated structure and the corresponding value for the optimal one. The values of Q$_{1}$, Q$_{2}$ and Q$_{3}$, as well as of $\theta$ and Q as a function of $\tilde{Q}_{1}$ are in Table ~\ref{tab:jtmodes}. 

\begin{table}[!h]
\caption[]{Values of Q$_{1}$, Q$_{2}$ and Q$_{3}$ as a function of $\tilde{Q}_{1}$ in plane 1 \& 2.}
\begin{tabular}{| c | c  c  c  c  c | c  c  c  c  c|}
\hline
$\tilde{Q}_{1}$	&  Q$_{1}$ (\AA) & Q$_{2}$ (\AA) & Q$_{3}$ (\AA) & $\theta$ (deg) & Q (\AA) &  Q$_{1}$ (\AA) & Q$_{2}$ (\AA) & Q$_{3}$ (\AA) & $\theta$ (deg) & Q (\AA)\\
\hline 
 & \multicolumn{5}{c |}{Plane 1 (0 0 0)} & \multicolumn{5}{c |}{Plane 2 (0 0 0.5)} \\
\hline
0   &	0	    &   0	    &   0	    &	  &      0 & 0   	    &   0	        &   0   	&	   & 0\\
0.25&	0.00789	&   0.00219	&   0.00041	&   79.35	&   0.00222 & 0.00789	&   -0.00219    &	0.00041	&   -79.35	&   0.00222\\
0.5	&   0.01579	&   0.00437	&   0.00082	&   79.35	&   0.00445 & 0.01579	&   -0.00437	&   0.00082	&   -79.35	&   0.00445\\
0.75&	0.02369	&   0.00656	&   0.00123	&   79.35	&   0.00667 & 0.02369	&   -0.00656	&   0.00123	&   -79.35	&   0.00667\\
1	&   0.03159	&   0.00875	&   0.00164	&   79.35	&   0.00890 & 0.03159	&   -0.00875	&   0.00164	&   -79.35	&   0.00890\\
\hline
\end{tabular}
\label{tab:jtmodes}
\end{table}

\subsection{Phase Diagram}

The phase diagram was derived by combining ab initio calculations and linear interpolation of the PSH parameters. 

The first-principles calculations were done by relaxing the structures with the cAFM configuration for U=2.8, 3.0, 3.2 and 3.4 eV.
Then, all the local structural parameters were linearly interpolated from the distorted to the undistorted (experimental-high-temperature) case. 
The parameters that are interpolated include the xy angles and the variation of the bond length which define the $\tilde{Q}_{1}$ parameter in Figure 3(c) of the main text. 

We have performed a linear interpolation of the fitting parameters,  by linearly interpolating the fitting coefficients from the distorted to the undistorted structure. Being a linear model suitable for the transition (See Figure 3(b) of the main text), we extrapolated the full phase diagram at higher values of $\tilde{Q}_{1}$.

\clearpage
\section{Band structure and optical spectra}

\begin{figure*}[h]
   \begin{center}
   \includegraphics[width=0.8\textwidth,clip=true]{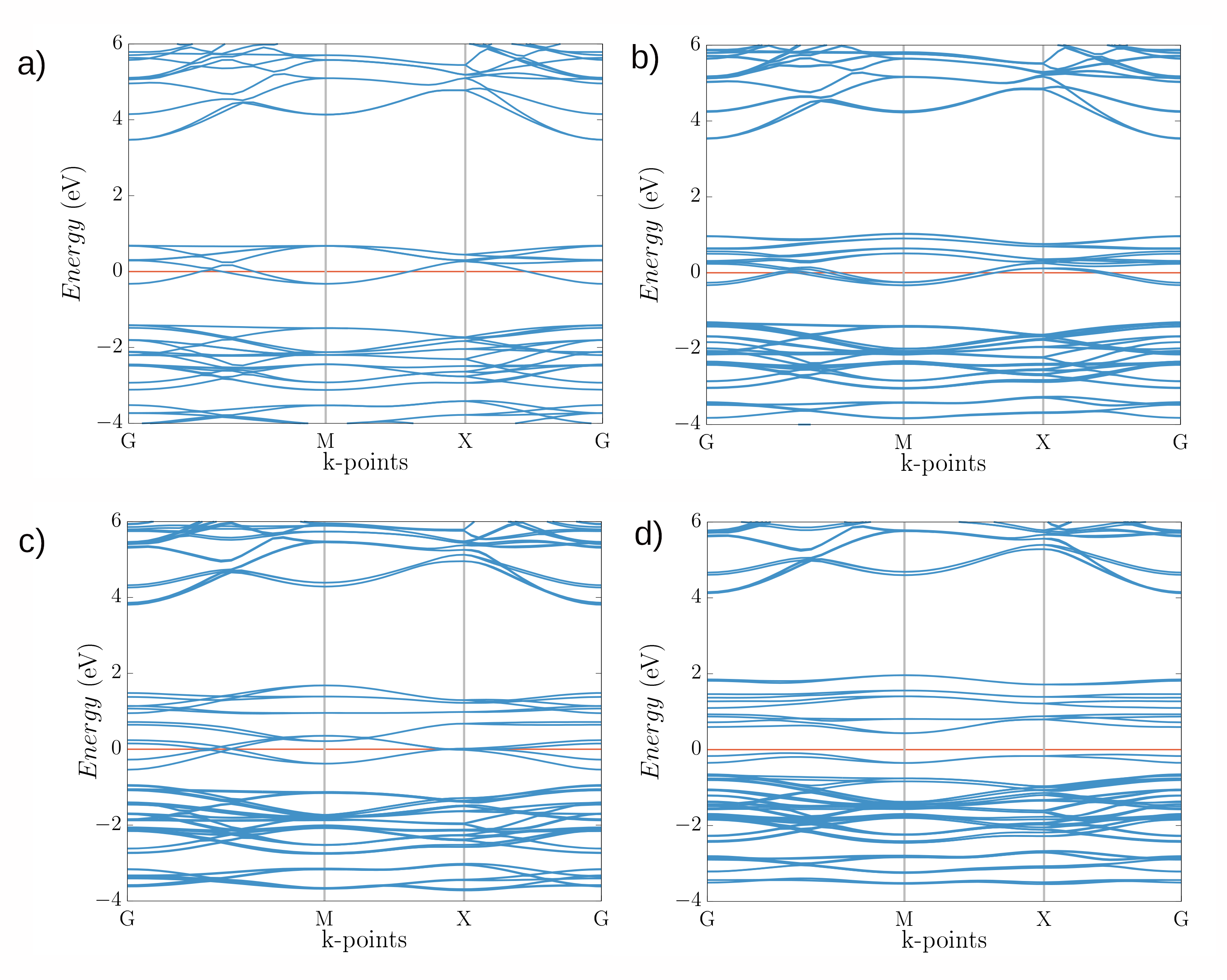}
    \end{center}
\caption{Band structure of BNOO with DFT (a), DFT+SOC (b), DFT+U (c) and DFT+U+ SOC (d). U is 3.4~eV.}
\label{fig:bands}
\end{figure*}

The Dirac-Mott nature of this compound is confirmed by our ab-initio calculations with DFT, DFT+SOC, DFT+U and DFT+U+SOC methods (see Figure ~\ref{fig:bands}).
Even with a high U of 5 eV we are not able to open the gap within the DFT+U scheme and only by including the relativistic effects  and a U larger than 1.5 eV, is possible to open the gap. 

\begin{figure*}[h!]
\centering\includegraphics[width=0.6\columnwidth,clip=true]{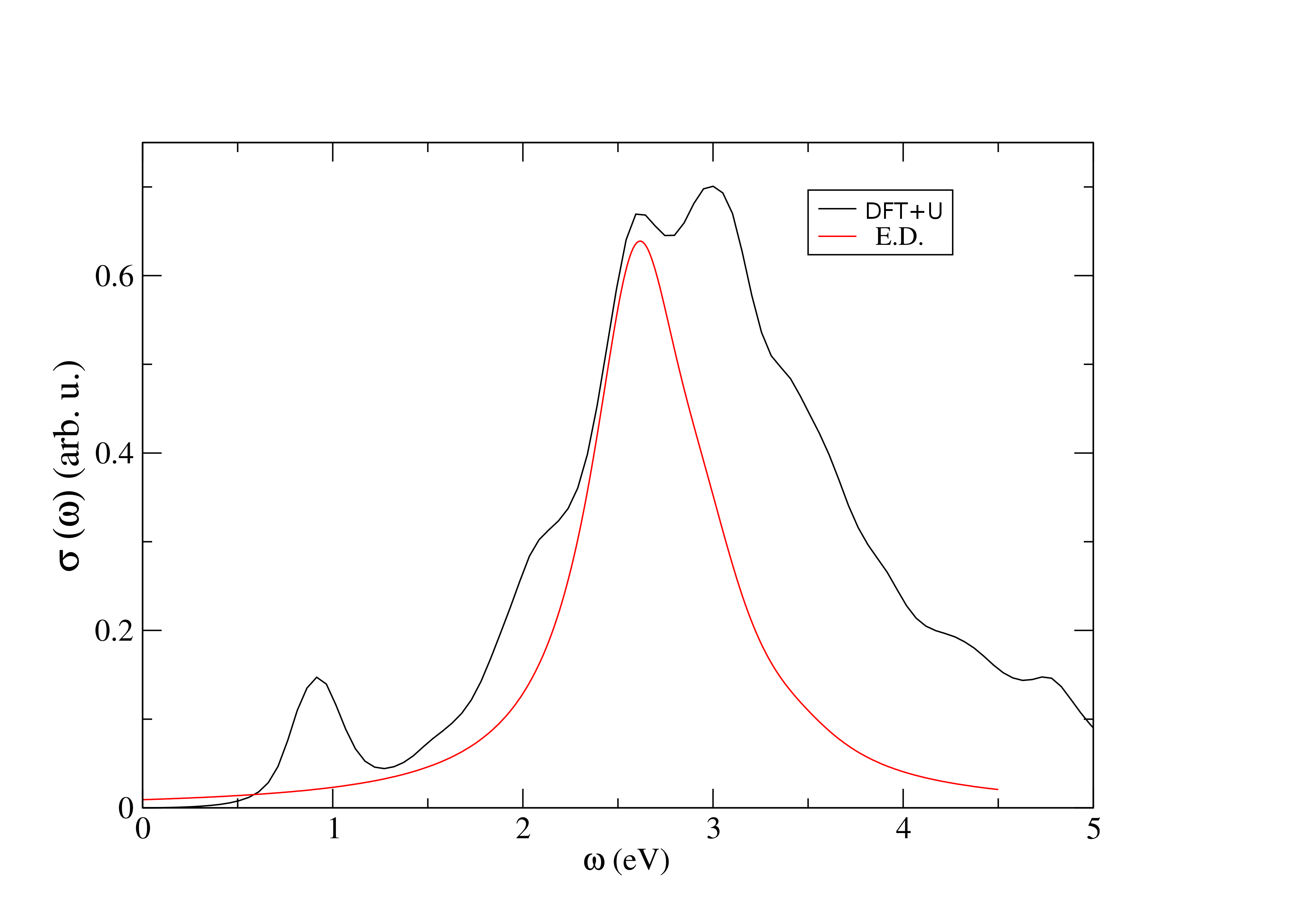}
\caption{DFT+U+SOC calculated optical spectra for Ba$_2$NAaOsO$_6$ compared with the ED prediction of the main Hubbard peak
(dominated by $t_{2g}-t_{2g}$ transitions.).}
\label{fig:spectra}
\end{figure*}

The band structure within DFT+U+SOC does not show significant changes upon different magnetic configurations, as well as by changing the local structure. 
The corresponding optical spectra at DFT+U*SOC level is shown in Fig.~\ref{fig:spectra}.

\section{Pseudo Spin Hamiltonian}

The exchange interactions were studied by mapping the DFT total energies onto Pseudo Spin Hamiltonian (PSH). 
The PSH used is given as a multipolar expansion of tensor products of dipolar, quadrupolar and octupolar operators that can be written, in its general form, as 
\begin{equation}
H_{ij} = \sum_{K,Q} I_{Q_{i},Q_{j}}^{K_{i},K_{j}} (i,j)O_{Q_{i}}^{K_{i}}(J)O_{Q_{j}}^{K_{j}}(J) \ ,
\end{equation}
where i,j label ionic sites, $K \leq 2J + 1$, $Q = -K, ..., K$ and $O_{Q}^{K}$ are multipolar tensor operators written in the spherical harmonic superbasis. The rank of such operators is given by K; where for K = 1 we have a  \emph{dipole}, K=2 identifies a  \emph{quadrupole} and K=3 the \emph{octupole}. 
The corresponding operators are written in Table ~\ref{tab:multipoles}.

\begin{table*}[b]
\caption[Multipolar operators for the BNOO Hamiltonian]{\textbf{Multipolar operators for the BNOO Hamiltonian.} The Multipole moments within a cubic  $\Gamma _{8}$ quartet. Bars over symbols correspond to permutations of dipole operators,  e.g. $\overline{j_{x}(j_{y})^{2}} = j_{x}j_{y}j_{y} + j_{y}j_{x}j_{y} + j_{y}j_{y}j_{x}$.}
\begin{tabular}{| c | c | c | c |}
\hline
Moment	& Spherical Superbasis & Moment & Spherical Superbasis  \\
\hline \hline
Dipole	& 	$O_{-1}^{1} = j_{y}$ & Octupole &$O_{-3}^{3} = \sqrt{5/8} \ (\overline{j^{2}_{x}j_{y}} - j_{y}^{3})$\\
	&	$O_{0}^{1} = j_{z}$	 & & $O_{-2}^{3} = \sqrt{5/12} \ \overline{j_{x}j_{y}j_{z}}$\\
	& $O_{1}^{1} = j_{x}$	& &	$O_{-1}^{3} = \sqrt{3/8} \ (4/3 \ \overline{j^{2}_{z}j_{y}} - 1/3 \ \overline{j^{2}_{x}j_{y}} - j_{y}^{3}) $	\\
Quadrupole	& $O_{-2}^{2} = \sqrt{3}/2 \ \overline{j_{x}j_{y}}$	 & & $O_{0}^{3} = j_{z}^{3} - 1/2 \ (\overline{j^{2}_{x}j_{z}} + \overline{j^{2}_{y}j_{z}})$	\\
		&$O_{-1}^{2} =  \sqrt{3}/2 \ \overline{j_{y}j_{z}}$	& &$O_{1}^{3} = \sqrt{3/8} \ (4/3 \ \overline{j^{2}_{z}j_{x}} - 1/3 \ \overline{j^{2}_{y}j_{x}} - j_{x}^{3}) $\\ 
		& $O_{0}^{2} = [3(j_{z})^{2} - j(j+1)]/2$  & &$O_{2}^{3} = \sqrt{5/12} \ (\overline{j^{2}_{x}j_{z}} - \overline{j^{2}_{y}j_{z}})$	\\
	&	$O_{1}^{2} = \sqrt{3}/2 \ \overline{j_{z}j_{x}}$ & & $O_{3}^{3} = \sqrt{5/8} \ (-\overline{j^{2}_{y}j_{x}} + j_{x}^{3})$	\\
		& $O_{2}^{2} =  \sqrt{3}/2 \ (j_{x}^{2} - j_{y}^{2})$  & &\\

		\hline
\end{tabular}
\label{tab:multipoles}
\end{table*}

\begin{figure*}[h]
   \begin{center}
   \includegraphics[width=0.2\textwidth,clip=true]{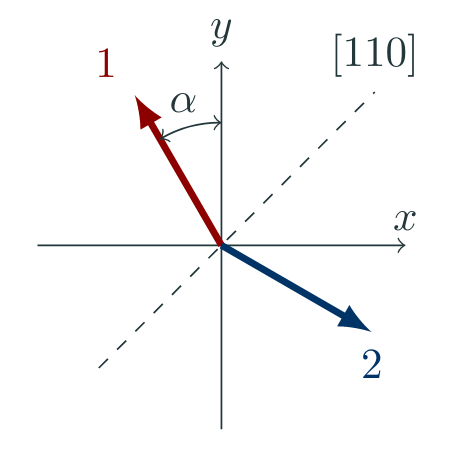}
    \end{center}
\caption{Global reference frame for the canting angles.}
\label{fig:new_reference_frame}
\end{figure*}

The PSH calculation proceeded in two steps. 
The first was to write the Pseudo Spin Matrices for a J = 3/2 Pauli pseudo spin and to project them onto the direction of the osmium magnetic moment. 
Due to the canting procedure described above, the coordinate system used is the global one of Figure ~\ref{fig:new_reference_frame}, and therefore, the unit vectors that describe the two osmium pseudo spins are given by
\begin{equation}
    \hat{n}_{1} = (sin(\alpha), cos(\alpha), 0) \ \ \ \ \ \ \  \hat{n}_{2} = (cos(\alpha), sin(\alpha), 0) \ .
\end{equation}
where $\alpha$ is defined in Figure ~\ref{fig:new_reference_frame} and it is related to $\phi$ via $\alpha = 45^{o} - \phi$.
Once this projection has been done, the new pseudo spin matrix was diagonalized and the eigenstate with highest eigenvalue was chosen. 

Then we applied the mean field analysis, summarized in equation $O(1)O(2) \approx \langle O(1)\rangle O(2)+ O(1)\langle O(2)\rangle - \langle O(1)\rangle\langle O(2)\rangle$, which allows to rewrite the total energy as a function of the mean values over pseudo eigenstates of the tensors operators, leading to
\begin{equation}
E = \frac{1}{2}\sum_{\langle i,j \rangle}\sum_{K,Q} I_{Q_{i},Q_{j}}^{K_{i},K_{j}} (i,j)  \langle O_{Q_{i}}^{K_{i}}(J) \rangle \langle O_{Q_{j}}^{K_{j}}(J) \rangle \ .
\label{eq:E_tot}
\end{equation}
From Table ~\ref{tab:multipoles} it is clear that every tensor contains powers of $j_{x}$, $j_{y}$ and $j_{z}$, and therefore, powers of the canting angle. 
Due to time reversal symmetry, only the even powers of pseudo spins are allowed, and, through algebraic simplifications, Eq.~\eqref{eq:E_tot} can be rewritten as 
\begin{equation}
E(\phi) = A_{0} +  \sum_{n=1}^{N} A_{n} cos(2n\phi) + \sum_{m = 1}^{M} B_{m} sin(2m\phi) \ . 
\label{eq:fittingcurve}
\end{equation}
Other restrictions come from the symmetry of the magnetic cell, which allows symmetric exchange-interaction matrices only in-plane (i.e. when Os(1) and Os(2) belong to the same plane), while anti-symmetric exchanges are allowed between out-of-plane interactions. 

The energy can be further separated into
\begin{equation}
E  = A_{0} + E_{d-d} + E_{q-q} + E_{d-o} + E_{o-o},
\label{eq:totham}
\end{equation}
where 
$E_{d-d}$ is the dipole-dipole contribution, $E_{q-q}$ is the quadrupole-quadrupole contribution,$E_{d-o}$ is the dipole-octupole contribution, $E_{o-o}$ is the octupole-octupole contribution. Converting Eq.~\eqref{eq:totham} into angular dependencies gives
\begin{equation}
E(\phi) = A_{0} +  \sum_{n=1}^{3} A_{n} cos(2n\phi) + \sum_{m = 1}^{2} B_{m} sin(2m\phi) \ , 
\label{eq:fittingcurve}
\end{equation}
where the coefficients $A_{n}$ are symmetric, while the coefficients $B_{m}$ come from broken inversion symmetry.


\subsection{Comparison DFT vs HI vs ED}

The comparison between DFT and HI/ED was achieved by reproducing the DFT+U+SOC canting curve starting from the calculated exchange coefficients obtained by HI and ED. This was done by reducing the  full 15 $\times$ 15 HI and ED interaction matrices to the relevant coefficients acting within the xy plane canting, based on a pseudo spin analysis. As previously specified, the mean values  $\langle O_{Q_{i}}^{K_{i}}(J) \rangle \langle O_{Q_{j}}^{K_{j}}(J) \rangle$ are powers of $cos(\phi)$ and $sin(\phi)$. These mean values were multiplied by the corresponding coefficients, collected and added, so to obtain the final expression with coefficients $A_{n}$ and $B_{m}$. 

As mentioned in the main text, in order to achieve a quantitatively consistent comparison between DFT+U and DFT+HI, the DFT+U+SOC AFM ($\phi$=90) - FM ($\phi=0$) energy difference has been aligned to the corresponding DFT+HI value. 
Specifically, the DFT+U+SOC curves (and therefore the corresponding magnetic interactions) have been divided by a factor of 10.  The ED results have not been rescaled, which guarantees a full consistency between DFT+HI and ED data, manifested by the very similar interaction matrices (see Tab.~\ref{tab:intermathi} and Tab.~\ref{tab:intermated})  and corresponding curves and exchange coefficients discussed in the main text. In the following we discuss the source of difference between DFT+U and effective Hamiltonian approaches and justify our choice to take DFT-HI as quantitative reference.

First we notice that (i) the two types of approaches employ different correlated basis set: PAW/FLAPW 'local' basis in VASP/Wien2k and $t_{2g}$ Wannier functions in DFT+HubI and ED;  moreover, (ii) while the DFT+U method calculates the total energies of competing magnetic phases, the ED and HubI-based approaches extract effective Hamiltonians by a perturbation-theory treatment of the local-moment paramagnetic phase.

Despite this fundamental difference between these two approaches, for localized $d$ system (in particular 3$d$ and to a lesser extent 4$d$) there is generally a good correspondence between these two basis and the results obtained are typically consistent, see for instance the results for SrTcO$_3$~\cite{PhysRevB.95.205115, PhysRevB.83.220402} and NiO~\cite{PhysRevB.84.115114,leonid}.
However, for more extended $d$ orbitals, where hybridization effects are generally large the degree of comparability between the two models is less robust. In effective Hamiltonian approaches the occupation of the orbital subspace is well defined and corresponds to the 'formal charge' (in our case one electron in the $t_{2g}$ orbitals), whereas in DFT the 'full occupation' is difficult to identify objectively as it depends on the choice of the integrated spheres and hybridization effects. This ambiguity between 'formal charge' and (DFT) 'full charge' is particularly evident in Os-double perovskites as discussed by Pickett for  BaCaOs$_2$O$_6$ (BCOO)~\cite{PhysRevB.93.155126}. In  BCOO the formal charge is 2, but DFT leads to a full occupation of about 5 electrons, due to the highly dispersive character of the Os 5$d$ orbital and the large degree of hybridization with the O-$p$ states. In our calculations we fall precisely in this situation as the full charge computed by VASP is 6 electrons, significantly larger than the corresponding formal charge considered in the model calculations (1 electron). 

Provided these significant differences in the correlated subspace occupancies between VASP and DFT+HubI/ED, the resulting large difference in the ordered state energetics is not  surprising. 
We would like to underline that 
formal and full $d$-shell occupation are virtually identical in NiO and (to a lesser extent) in SrTcO$_3$; in the latter DFT+U occupancy is 3.43 as compared to the formal 4$d$ occupancy of 3 \cite{PhysRevB.95.205115}. The similarity of formal and full $d$-shell occupancies  in those cases results in the exchange energies being very similar in the both approaches. 
Conversely, for the 5$d$ oxide UO$_2$ quantitative difference between DFT+U+SOC and HI were already reported in literature~\cite{PhysRevMaterials.3.083802, PhysRevB.99.094439}.

Support on the quantitative validity of the effective Hamiltonian approaches can be  inferred from the comparison of calculated and experimental N\'eel temperature $T_N$. The mean-field value of $T_N$=17.5~K obtained from the calculated FT-HI interactions for the undistorted phase is in a reasonable agreement, taking into account the usual mean-field overestimation, with its experimental value of 6.7~K.  For lattice distortions to occur at the AFM transition, the ordering energy for the distorted structure must be larger than that for the undistorted one. Hence, one would expect the ordering energy (and, similarly, $T_N$) for the distorted phase to be at least not smaller than that of the undistorted one. The ordering energy and (to a lesser extent) transition temperature of the undistorted case thus provide a lower limit for those quantitites in the distorted case. Since undistorted FT-HI $T_N$ already has a  correct (and even  somewhat overestimated) magnitude, one would not expect the magnetic energy to be significantly larger than that given by FT-HI (and ED).   

From these arguments the overall magnitude of FT-HI/ED interactions seems to be in a better correspondence to experiment as compared to DFT+U, hence, we normalized the magnitude of DFT+U interactions accordingly. However, notice that all three approaches predict virtually the same non-trivial shape of magnetic energy vs. angle $\phi$ (Fig. 2b). Correspondingly, the relative magnitudes of  coefficients  $A_1$, $A_2$ and $A_3$  in the undistorted phase also agree well. Basing on these observations one may be confident that the qualitative shape of magnetic energy vs. $\phi$ and the relative magnitude of various multipolar contributions into it are correctly resolved by the DFT+U method.

\clearpage
\section{Appendix}

CIF file created by FINDSYM, version 7.0\\
data\_findsym-output\\
\_audit\_creation\_method FINDSYM\\

\_cell\_length\_a    5.8597900000\\
\_cell\_length\_b    5.8597900000\\
\_cell\_length\_c    8.2870000000\\
\_cell\_angle\_alpha 90.0000000000\\
\_cell\_angle\_beta  90.0000000000\\
\_cell\_angle\_gamma 90.0000000000\\
\_cell\_volume      284.5518696011\\

\_symmetry\_space\_group\_name\_H-M "P 21/n 21/n 2/m"\\
\_symmetry\_Int\_Tables\_number 58\\
\_space\_group.reference\_setting '058:-P 2 2n'\\
\_space\_group.transform\_Pp\_abc a,b,c;0,0,0

loop\_\\
\_space\_group\_symop\_id\\
\_space\_group\_symop\_operation\_xyz\\
1 x,y,z\\
2 x+1/2,-y+1/2,-z+1/2\\
3 -x+1/2,y+1/2,-z+1/2\\
4 -x,-y,z\\
5 -x,-y,-z\\
6 -x+1/2,y+1/2,z+1/2\\
7 x+1/2,-y+1/2,z+1/2\\
8 x,y,-z\\

loop\_\\
\_atom\_site\_label\\

\_atom\_site\_type\_symbol\\
\_atom\_site\_symmetry\_multiplicity\\
\_atom\_site\_Wyckoff\_label\\
\_atom\_site\_fract\_x\\
\_atom\_site\_fract\_y\\
\_atom\_site\_fract\_z\\
\_atom\_site\_occupancy\\
\_atom\_site\_fract\_symmform\\
Na1 Na   2 b 0.00000 0.00000 0.50000 1.00000 0,0,0\\   
Os1 Os   2 a 0.00000 0.00000 0.00000 1.00000 0,0,0  \\ 
Ba1 Ba   4 f 0.00000 0.50000 0.25000 1.00000 0,0,Dz  \\
O1  O    4 g 0.77126 0.23104 0.00000 1.00000 Dx,Dy,0 \\
O2  O    4 g 0.22667 0.22888 0.00000 1.00000 Dx,Dy,0 \\
O3  O    4 e 0.00000 0.00000 0.22849 1.00000 0,0,Dz  \\


\end{document}